\documentclass{article}
\usepackage{ijcai22}

\usepackage{times}
\usepackage{soul}
\usepackage{url}
\usepackage[hidelinks]{hyperref}
\usepackage[utf8]{inputenc}
\usepackage[small]{caption}
\usepackage{graphicx}
\usepackage{amsmath}
\usepackage{amsthm}
\usepackage{booktabs}
\usepackage{algorithm}
\usepackage{algorithmic}
\usepackage{subfig}
\usepackage{bm}
\usepackage{multirow}
\usepackage{bigstrut}

\title{SyntaSpeech: Syntax-Aware Generative Adversarial Text-to-Speech}

\author{
	Author Name
	\affiliations
	Affiliation
	\emails
	pcchair@ijcai-22.org
}

\author{
	Zhenhui Ye$^{1}$
	\and
	Zhou Zhao$^{1}$
	\and
	Yi Ren$^{1}$
	\and
	Fei Wu$^{1}$
	\affiliations
	$^1$College of Computer Science and Technology, Zhejiang University\\
	\emails
	\{zhenhuiye, zhaozhou, rayeren, wufei\}@zju.edu.cn
}

\begin{document}
	
	\maketitle
	
	\begin{abstract}
		The recent progress in non-autoregressive text-to-speech (NAR-TTS) has made fast and high-quality speech synthesis possible. However, current NAR-TTS models usually use phoneme sequence as input and thus cannot understand the tree-structured syntactic information of the input sequence, which hurts the prosody modeling. To this end, we propose SyntaSpeech, a syntax-aware and light-weight NAR-TTS model, which integrates tree-structured syntactic information into the prosody modeling modules in PortaSpeech \cite{ren2021portaspeech}. Specifically, 1) We build a syntactic graph based on the dependency tree of the input sentence, then process the text encoding with a syntactic graph encoder to extract the syntactic information. 2) We incorporate the extracted syntactic encoding with PortaSpeech to improve the prosody prediction. 3) We introduce a multi-length discriminator to replace the flow-based post-net in PortaSpeech, which simplifies the training pipeline and improves the inference speed, while keeping the naturalness of the generated audio. Experiments on three datasets not only show that the tree-structured syntactic information grants SyntaSpeech the ability to synthesize better audio with expressive prosody, but also demonstrate the generalization ability of SyntaSpeech to adapt to multiple languages and multi-speaker text-to-speech. Ablation studies demonstrate the necessity of each component in SyntaSpeech\footnote{Source code and audio samples are available at \url{https://syntaspeech.github.io}, }.          
	\end{abstract}
	
	\section{Introduction}
	Text-to-speech (TTS) aims to synthesize natural speech for input text. Recently, deep learning based TTS has made rapid progress and shown competitive performance with traditional TTS systems \cite{oord2016wavenet}. Neural TTS approaches typically learn an acoustic model that generates the mel-spectrogram or linguistic features from the input sentence \cite{wang2017tacotron}, then adopt a vocoder to synthesize the waveform \cite{oord2016wavenet}. To effectively extract semantic and prosody information from the input text, some previous neural TTS models generate mel-spectrograms autoregressively and suffer from slow inference speed \cite{ping2018deepvoice3}. To improve the practicality, non-autoregressive text-to-speech (NAR-TTS) explores to synthesize the mel-spectrogram in parallel \cite{ren2019fastspeech}, yet is faced with the difficulty to model expressive prosody using non-autoregressive structures. Recently, NAR-TTS modules tackle this problem by decoupling the prosody into several aspects (such as duration, pitch, etc) \cite{kim2020glowtts}\cite{ren2020fastspeech2}, and achieves comparable performance with autoregressive text-to-speech approaches (AR-TTS). Currently, improving the modeling of the prosody is still an open question in NAR-TTS.
	
\begin{figure}[!t]	
		\centering
		\includegraphics[width=7.5cm]{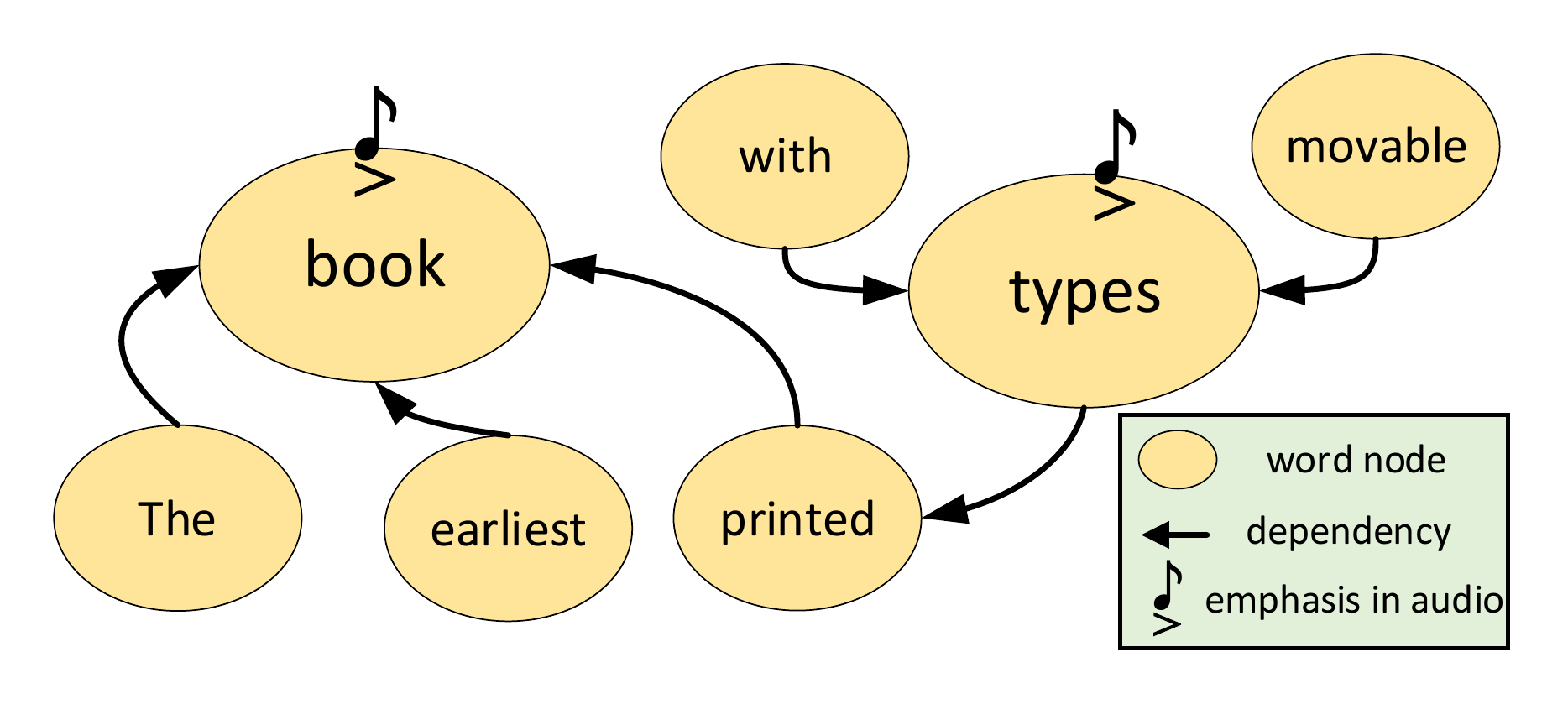}
		\caption{The dependency tree of the input text "\textit{The earliest book printed with movable types}". The emphasis in the real audio is marked with the emphasis symbol.}
		\label{figure:english_dependency_example}	
	\end{figure}

	Syntactic information, especially the dependency relation, possesses rich intonational features such as pitch accent and phrasing of the input text \cite{hirschberg2001prosody-and-syntax}. To be intuitive, we provide an example in Fig.\ref{figure:english_dependency_example} to show the potential relationship between the dependency tree and the audio. There are also many TTS extensions utilizing syntactic information to improve prosody. For instance, GraphTTS \cite{sun2020graphtts} and GraphPB \cite{sun2021graphpb} construct a syntactic graph based on the character sequence and prosody boundary in the sentence, respectively. GraphSpeech \cite{liu2021graphspeech} and RGNN \cite{zhou2021rgnn} utilize dependency relation in a sentence and extract the syntactic information with graph neural networks. However, previous syntax-aware TTS models are done in the framework of AR-TTS. Since AR-TTS predicts the duration and pitch autoregressively, it could easily exploit the syntactic information by taking it as auxiliary input features of the backbone. By contrast, NAR-TTS typically models prosody with external predictors, although the extracted features can be used as the auxiliary input features of these
	prosody predictors, this approach has not been explored yet. To our knowledge, there is no NAR-TTS model that could effectively embed the tree-structured syntactic information to improve the prosody prediction.
	
	To exploit the syntactic information with NAR-TTS, in this work, we propose SyntaSpeech, a syntax-aware generative text-to-speech model, which improves the prosody in the generated mel-spectrogram using a graph encoder to exploit the dependency relation of the raw text, and enhances the audio quality with adversarial training. Specifically,
	\begin{itemize}
		\item To generate the word-level syntactic encoding, we build a syntax graph for each input sentence based on its dependency tree, process the phoneme-level latent encoding to represent the word node in the graph, then aggregate the graphical features with a graph encoder.
		\item To utilize the extracted syntactic features in prosody modeling, we incorporate the graph encoder into PortaSpeech. The syntactic encoding is embedded into the duration predictor and the variational generator, to improve the duration and pitch prediction, respectively. 
		\item To generate realistic audio with lightweight structures and simplify the training pipeline, we adopt multi-length adversarial training to replace the flow-based post-net in PortaSpeech.
	\end{itemize}
	
	To demonstrate the generalization ability of our SyntaSpeech, we perform experiments on three datasets, including one single-speaker English dataset, one single-speaker Chinese corpus, and one multi-speaker English dataset. Experiments on all datasets show that SyntaSpeech outperforms other state-of-the-art TTS models in voice quality and (especially) prosody in terms of subjective and objective evaluation metrics. The rest of the paper is organized as follows: In Sec.\ref{sec:related_works} we discuss recent progress in NAR-TTS and previous works that develop a syntax-aware TTS model. In Sec.\ref{sec:syntaspeech} we introduce our SyntaSpeech in details. Performance evaluation and ablation studies of SyntaSpeech are given in Sec.\ref{sec:experiments}. Finally, we draw conclusions in Sec.\ref{sec:conclusion}.
	
	\section{Related Works}
	\label{sec:related_works}
	\subsection{Non-Autoregressive Text-to-Speech}
	In the past few years, modern neural TTS thrived with the development of deep learning. Originally, to model the long-term relationships among the input tokens, previous works tend to generate the mel-spectrogram autoregressively \cite{wang2017tacotron}\cite{ping2018deepvoice3}. However, AR-TTS is faced with the
	challenges of slow inference and robustness issues (e.g., word skipping) incurred by autoregressive generation.
	
	To tackle these issues, many works explore adopting non-autoregressive generation. Some works use positional attention for the text and speech alignment\cite{peng2020non}, while the other works use duration prediction to handle the length mismatch between text and mel-frame sequences. For instance, FastSpeech \cite{ren2019fastspeech}, Glow-TTS \cite{kim2020glowtts}, and EATS \cite{donahue2020eats} use duration predictor to upsample the phoneme sequence to match the length of mel-spectrograms. These works enjoy fast inference and well robustness. Recent works further improve the expressiveness in NAR-TTS by modeling the variation information. For instance, FastSpeech 2 \cite{ren2020fastspeech2} introduced a pitch predictor to infer the pitch contour in the generated mel-spectrogram. VITS \cite{kim2021vits} and PortaSpeech \cite{ren2021portaspeech} leverage variational auto-encoder (VAE) to model the variation information in the latent space. To date, improving the expressiveness of the generated waveform is still an open question to the TTS community. 
	
	\subsection{Syntax-aware Text-to-Speech}
	Syntax information, which records the dependency relation between the tokens in the text, is acknowledged as a helpful feature to estimate the prosody of the speech and has been studied in speech synthesis before the neural TTS age \cite{hirschberg2001prosody-and-syntax}\cite{mishra2015intonational}.
	
	Modern TTS typically utilizes the syntactic information as auxiliary features in AR-TTS modules: GraphTTS \cite{sun2020graphtts} designs a character-level text-to-graph module to extract the sequential information in the sentence and tries several graph neural networks (GNNs) to process the graphical features. The extracted syntactic feature is then fed into the decoder of Tacotron \cite{wang2017tacotron} as an auxiliary encoding. Later, GraphSpeech \cite{liu2021graphspeech} introduces \textit{dependency parsing} in the text-to-graph module to better represent the syntactic information of the input sentence, and utilizes bi-directional gated recurrent unit (GRU) \cite{cho2014gru} to aggregate information through the syntactic graph. Recently, RGGN \cite{zhou2021rgnn} also adopts dependency parsing to construct the syntactic graph and utilize pre-trained word embedding from BERT \cite{devlin2019bert}, then process the graphical data with gated graph neural network (GGNN) \cite{li2015gated}. Both of GraphSpeech and RGGN regard the dependency-based syntactic encoding as auxiliary features and feed them into the encoder of the sequence-to-sequence (seq-to-seq) AR-TTS module.
	
	The difference between SyntaSpeech and previous works is as follows. Firstly, to our knowledge, our SyntaSpeech is the first work that analyzes the syntactic information in NAR-TTS. Secondly, previous works extract syntactic information to provide a better text representation for the seq-to-seq model, while we learn the syntactic encoding for the duration and other prosody attributes prediction, which could make full use of the syntactic features and is more interpretable. Thirdly, previous works either use pre-trained embedding or learn character-level embedding as the node representation in the syntactic graph, by contrast, we process the latent features in the backbone of the TTS model with word-level pooling \cite{ren2021portaspeech} to formulate the node embedding.
	
	\section{SyntaSpeech}
	\label{sec:syntaspeech}
	To exploit the syntactic information of the input text in the framework of NAR-TTS, we propose SyntaSpeech, which exploits the dependency relation to improve the naturalness and expressiveness of the synthesized audio waveform. In this section, we first introduce a \textit{syntactic graph builder} to construct a syntactic graph based on the input text, which can be utilized in either English or Chinese. Then we design the overall network structure of SyntaSpeech based on PortaSpeech \cite{ren2021portaspeech}. As shown in Figure 1a, SyntaSpeech designs a \textit{syntactic graph encoder} to provide syntactic information for duration prediction (in linguistic encoder) and other prosody attributes distribution modeling (in variational generator). In general, SyntaSpeech exploits the syntactic information in the raw text with the following steps: 
	\begin{itemize}
		\item Firstly, the text sequence is fed into the transformer-based phoneme encoder to obtain the phoneme encoding, which is then processed into a word-level representation with average pooling based on the word boundary.
		\item Secondly, the syntactic graph builder constructs the syntactic graph using dependency relation, and the word encoding is aggregated through the constructed graph using \textit{gated graph convolution} \cite{li2015gated}.
		\item Thirdly, the obtained word-level syntactic encoding is expanded into phoneme level and frame level, to embed syntactic information into the duration prediction and pitch-energy prediction, respectively.
	\end{itemize}
	
	Besides, we also replace the post-net in PortaSpeech with adversarial training to simplify the training pipeline while keeping the naturalness of the generated mel-spectrogram. We describe these designs in detail in the following subsections. More technical details are provided in Appendix A.
	\begin{figure*}[!t]
		\centering
		\subfloat[SyntaSpeech]{\includegraphics[scale=0.43]{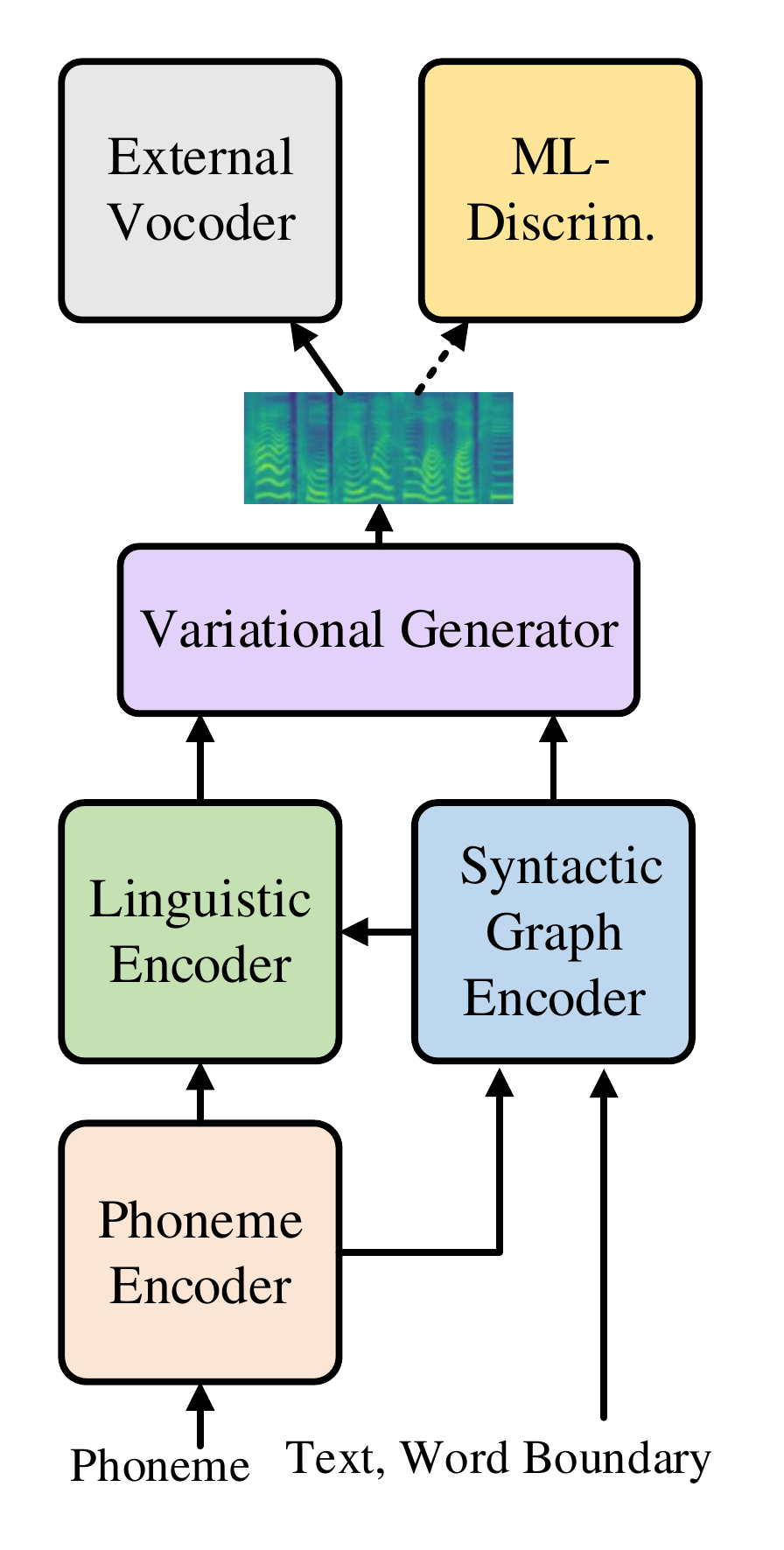}%
			\label{figure:overall_structure}}
		\hfil
		\subfloat[Syntactic Graph Encoder]{\includegraphics[scale=0.43]{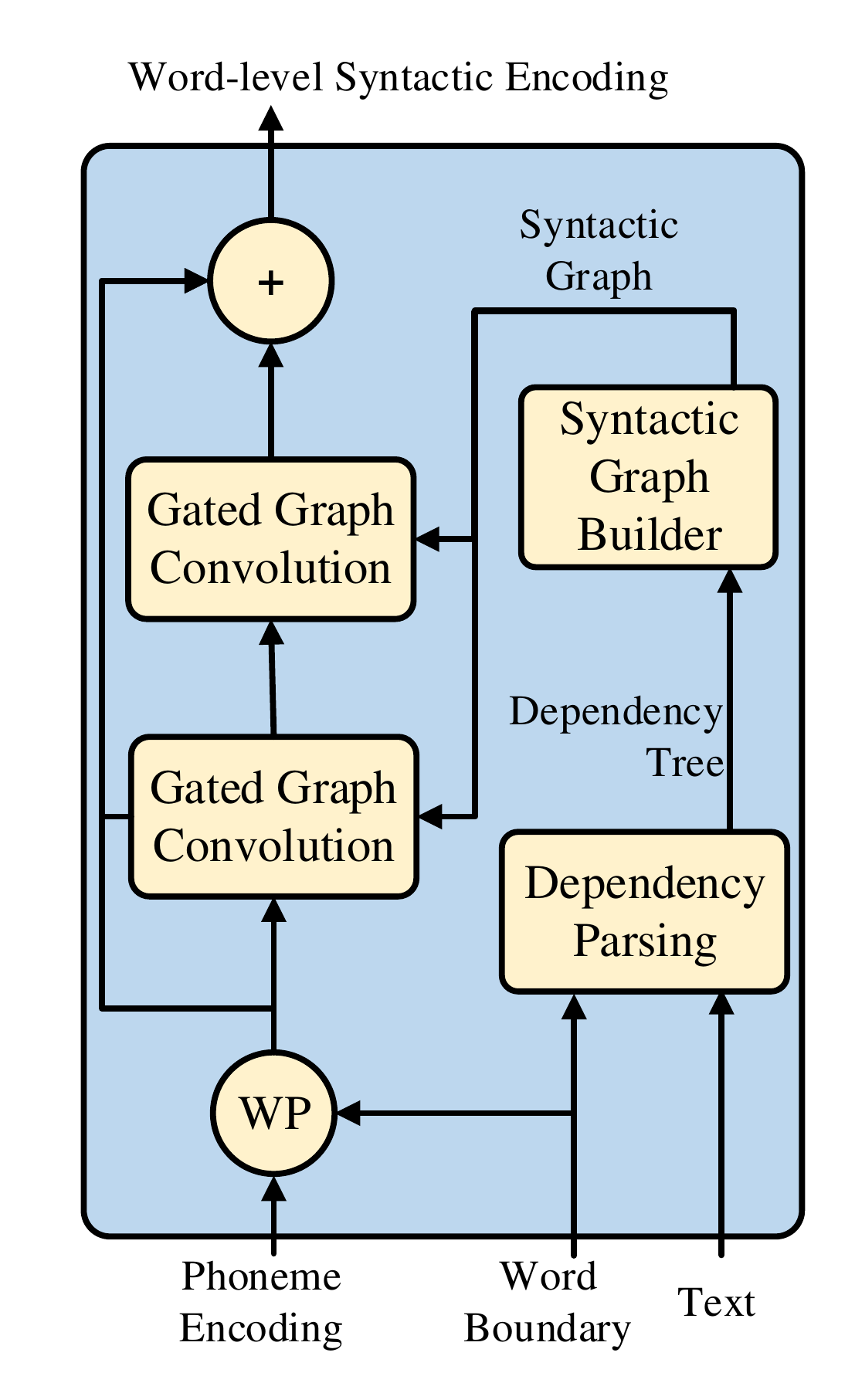}%
			\label{figure:graph_encoder}}
		\hfil
		\subfloat[Linguistic Encoder]{\includegraphics[scale=0.425]{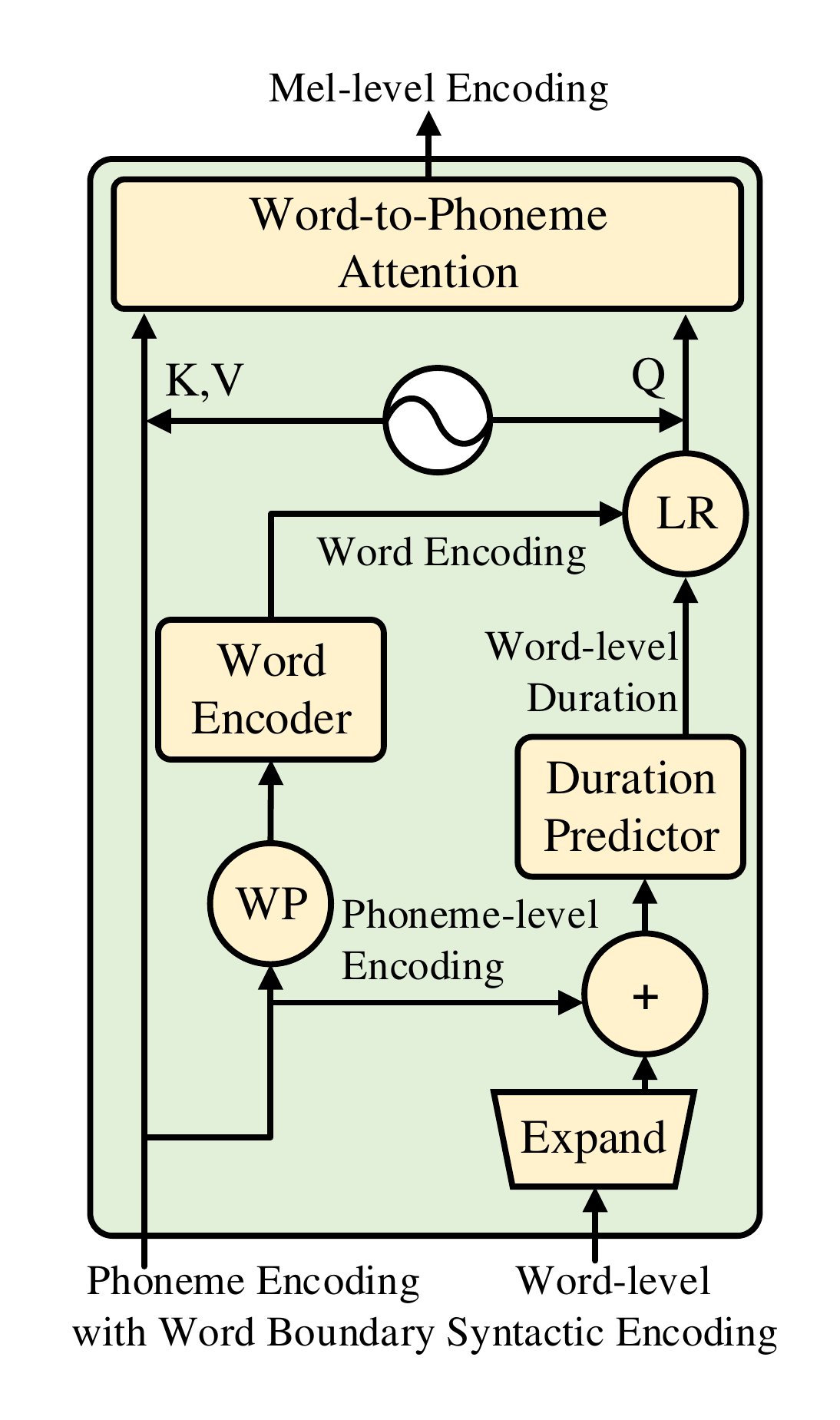}%
			\label{figure:linguistic_encoder}}
		\hfil
		\subfloat[Variational Generator]{\includegraphics[scale=0.435]{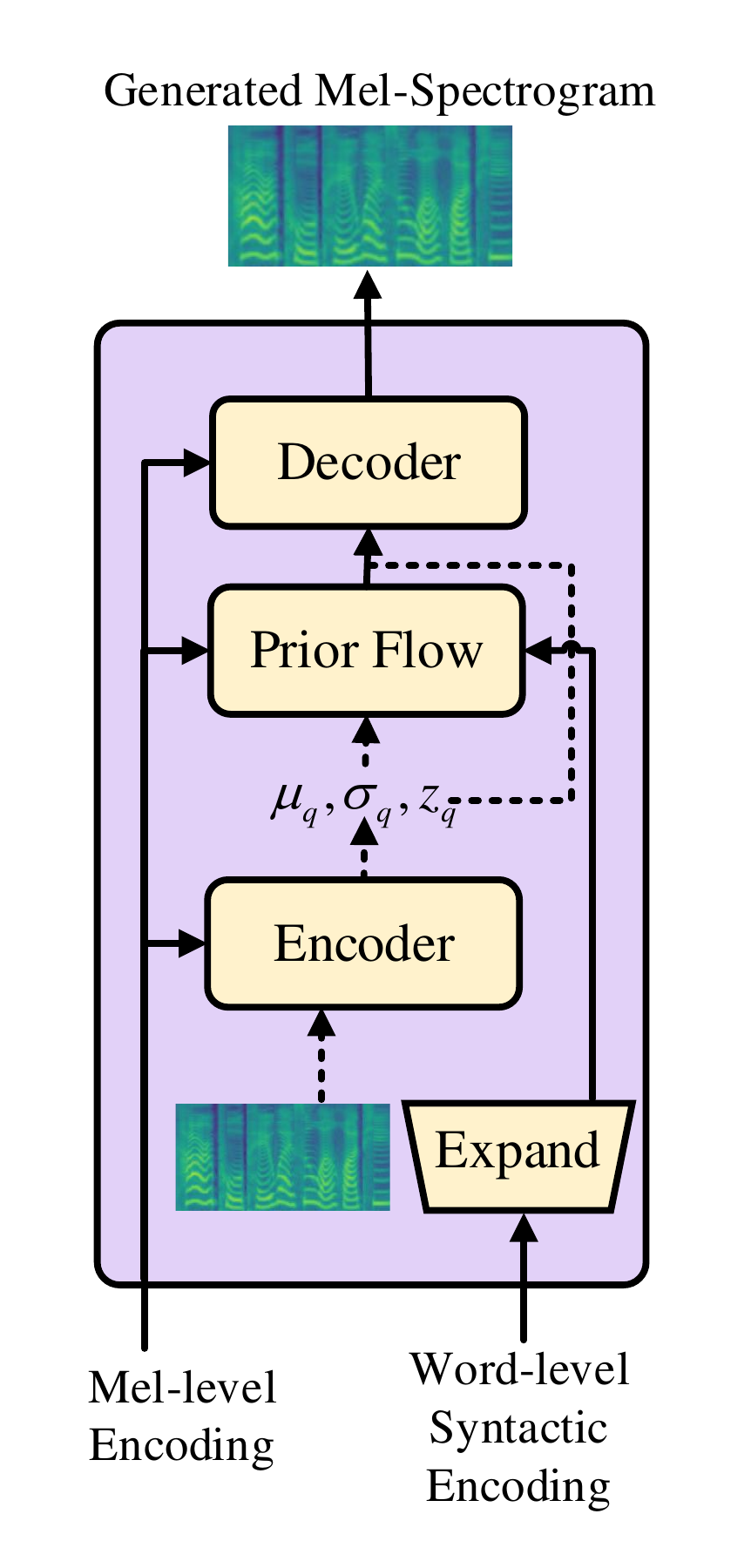}%
			\label{figure:variational_generator}}
		\caption{The overall structure for SyntaSpeech. In subfigure(a), "ML-Discrim" denotes Multi-Length Discriminator in HiFiSinger. In subfigure (b), "WP" denotes the word-level average pooling operation, and the "Syntactic Graph Builder" is illustrated in Sec.\ref{sec:graph_builder}. In subfigure (c), "LR" denotes the Length Regulator proposed in PortaSpeech. In subfigures (a) and (d), the dashed lines denote that the operations are only executed in the training phase.}
		\label{figure:network_structure}	
	\end{figure*}

	\subsection{Syntactic Graph based on Dependency Relation}
	\label{sec:graph_builder}
	Dependency parse tree can be regarded as a directed graph, where each edge represents the dependency relation between two nodes (words). It provides a hierarchical representation for plain text sentences and is considered to contain rich syntactic information. To make full use of the syntactic information contained in the dependency tree, we introduce a \textit{syntactic graph builder} to convert the dependency tree (or say the raw dependency graph) into a \textit{syntactic graph}, which is more compatible with graph neural networks and existing NAR-TTS structures.
	
	The biggest challenge in extracting syntactic information with GNNs is the single-directed structure of the raw dependency graph, which denotes that the leaf node in the graph cannot obtain any information from other nodes during the graph aggregation. To handle this, inspired by previous works that exploit dependency relation in AR-TTS, we add a reverse edge for each directed edge in the dependency tree so that the information flow in the graph is bi-directional. Specifically, there could be \textit{forward edges} from parent nodes to child nodes, which is consistent with the dependency tree, as well as \textit{reversed edges} from child nodes to parent nodes. Then, we introduce our methods of constructing syntactic graphs with node embedding in specific languages.
	
\begin{figure}[!t]
		\centering
		\subfloat[The graph built from the dependency tree in Fig.\ref{figure:english_dependency_example}.]{\includegraphics[width=7cm]{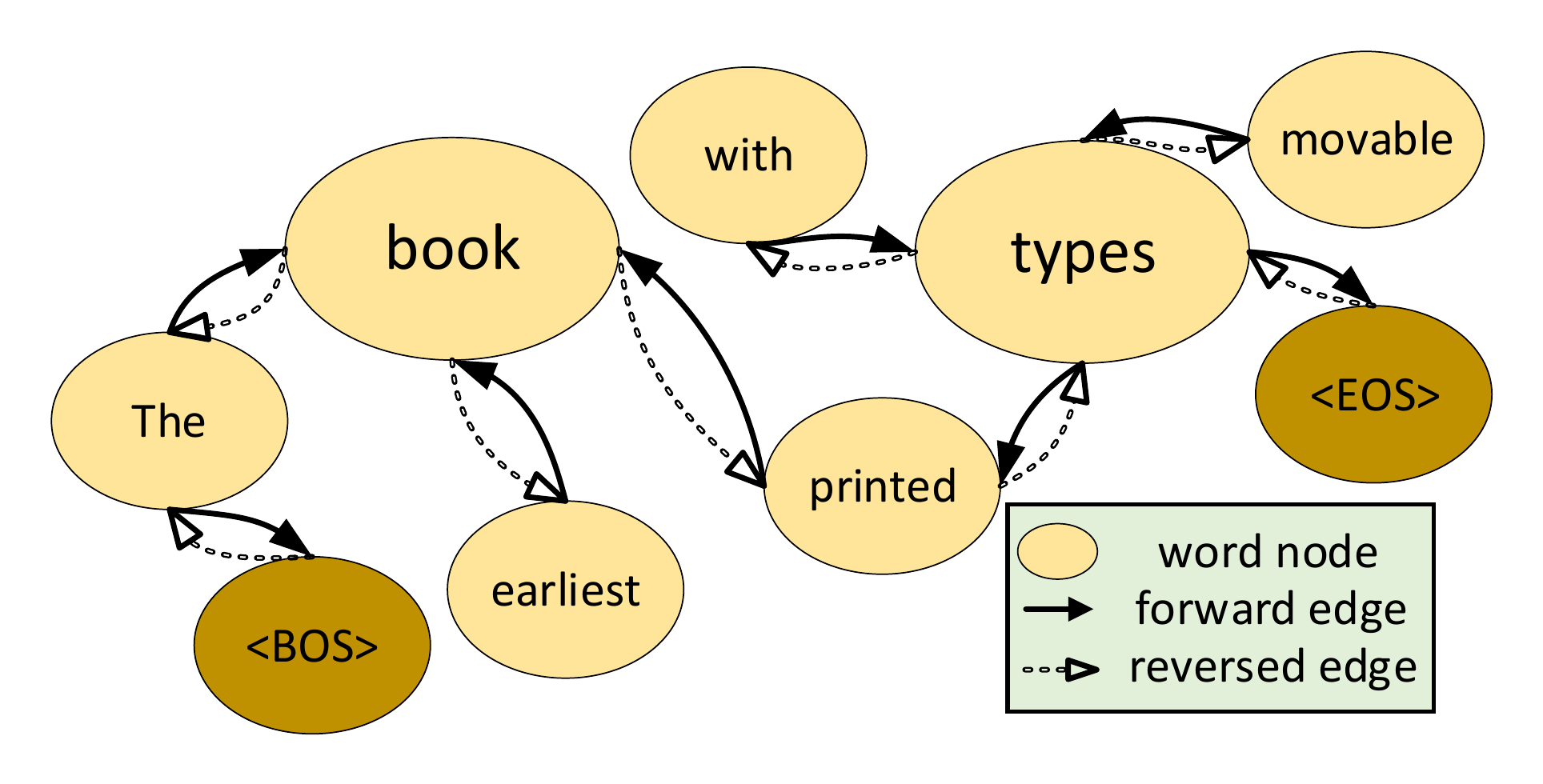}%
			\label{figure:english_graph_example}}
		\hfil
		\subfloat[The graph built from a Chinese dependency tree.]{\includegraphics[width=7.5cm]{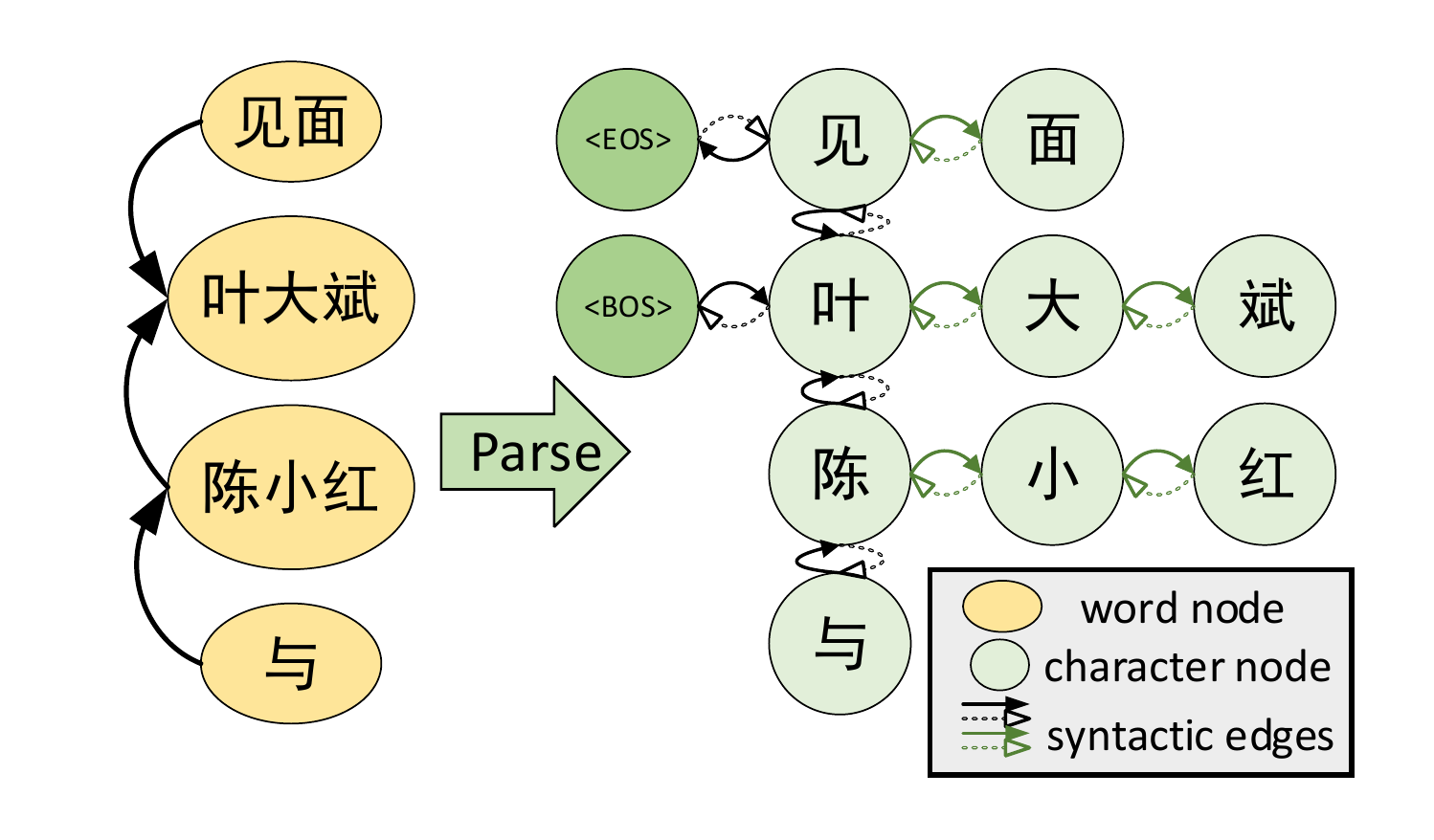}%
			\label{figure:chinese_example}}
		\caption{Two examples of syntactic graph construction.}
		\label{figure:build_graph_examples}
	\end{figure}

	\paragraph{Graph for English} To construct the syntactic graph for English text, we add \textit{BOS} and \textit{EOS} into the above-mentioned bi-directional graph and connect them with the first and last words of the input sentence, respectively. To be intuitive, we provide an example that transforms an English sentence into syntactic graph in Fig.\ref{figure:english_graph_example}, where \textit{forward edges} are represented as solid black arrows and \textit{reversed edges} are dashed black arrows. Then we consider the node representation in the constructed syntactic graph. Note that while TTS models typically use phoneme sequence as the input, the word is the fundamental unit in dependency parsing. To obtain the word-level node embedding, inspired by PortaSpeech, we adopt word-level average pooling to the phoneme encoding with word boundary information to generate the word encoding. As our node embedding is the latent encoding in the TTS model, it possesses valuable acoustic features for the TTS task and can be jointly optimized through backpropagation.
	
	\paragraph{Graph for Chinese } As for the Chinese dataset, we make small adaptations. Different from English where the pronunciation of the word is directly decided by the phoneme, in Chinese the phoneme decides the pronunciation of the Chinese character, and the character decides the pronunciation of the word. To make the node representation more compatible with the Chinese pronunciation law, instead of extracting the word-level encoding as we design for English, we adopt character-level average pooling to generate the character encoding. To be coherent to the obtained character encoding, we extend the \textit{syntactic graph} by expanding each word node into several Chinese character nodes, then use the first character node in each word to make the inter-word dependency connection, and other characters are sequentially connected according to the order in the word. Therefore, we additionally define two edges to represent the intra-word connection in \textit{forward} and \textit{reversed} directions, respectively. An intuitive example is shown in Fig.\ref{figure:chinese_example}, where the green solid/dashed arrows denote the intra-word forward/reversed edges.
	
	\paragraph{Graph for other languages} The syntactic graph for other languages can be constructed similarly. For instance, French and Spanish datasets can directly follow our approach for English, while Japanese datasets can use our graph construction method for Chinese.
	
	\subsection{Syntax-Aware Graph Encoder for Prosody Prediction}
	To learn the syntax-aware word representation from the input text, we design a \textit{syntactic graph encoder} based on the \textit{syntactic graph builder} and GNNs, which is shown in Fig. \ref{figure:graph_encoder}. As illustrated in Sec.\ref{sec:graph_builder}, we process the input text with word boundary with the syntactic graph builder to generate a syntactic graph with heterogeneous edges (2 edges for English and 4 edges for Chinese), and in the meantime, the phoneme embedding is processed with word-level average pooling to formulate the node embedding in the syntactic graph. Now that the syntactic graph is equipped with learnable node embedding, the syntactic information is extracted through graph aggregation as follows: 1) we utilize two stacked Gated Graph Convolution layers with both 5 iterations to extract the long-term dependency in the graph; 2) the output of all preceding layers are summed up as the output syntactic word-level encoding, so as to assemble and reuse the word-level features from different receptive fields in the syntactic graph.
	
	Then we consider embedding the extracted syntactic word encoding into the TTS model. SyntaSpeech keeps main structures of PortaSpeech: a Transformer-based \textit{linguistic encoder} to extract frame-level semantic representations with the help of a word-level duration predictor;  a VAE-based \textit{variational generator} with flow-based prior to synthesize the predicted mel-spectrogram. With these structures, PortaSpeech divides the prosody prediction (including duration, pitch, energy, etc.) into two sub-tasks: the duration predictor in linguistic encoder controls the timing in word-level; and in the variational generator, a flow-based enhanced prior distribution is introduced to predict the pitch, energy, and other prosody attributes. Based on the above insights, SyntaSpeech learns two individual syntactic graph encoders to extract syntactic features for duration prediction and 
	other prosody attributes (e.g., energy and pitch) distribution modeling, respectively. To be specific, the extracted syntactic word encoding of the first graph encoder is expanded into phoneme-level and be fed into the duration predictor (as shown in Fig.\ref{figure:linguistic_encoder}), and the output of the second graph encoder is expanded into frame level as auxiliary features of the prior flow in variational generator (as shown in Fig. \ref{figure:variational_generator}). 
	
	\subsection{Multi-Length Adversarial Training}
	
	The mel-spectrogram prediction of TTS models learned with mean square error (MSE) or mean absolute error (MAE) is generally challenged with blurry outputs. To handle this, PortaSpeech introduces a flow-based post-net to refine the predicted mel-spectrogram of the variational generator. Another common practice in handling the over-smoothing problem is to adopt the adversarial loss \cite{binkowski2019gan-tts}\cite{donahue2020eats}. Following HiFiSinger \cite{chen2020hifisinger}, we introduce a \textit{multi-length discriminator} to distinguish between the output generated by the TTS model and the ground truth mel-spectrogram. Specifically, the variational generator is coupled with an ensemble of multiple CNN-based discriminators which evaluates the generated (true) spectrogram based on random windows of different lengths. Detailed structures can be found in Appendix A.2. Compared with using post-net in PortaSpeech, the benefits of multi-length adversarial training are twofold: 1) it can generate realistic spectrogram similar to post-net yet at a faster inference speed; 2) it can better capture unnatural slice in the generated sample and help improve the naturalness of word pronunciation. 
	
	\section{Experiments}
	\label{sec:experiments}
	\subsection{Experimental Setup}
	\label{sec:experimental_setup}
	\paragraph{Datasets and Baselines}  We evaluate SyntaSpeech on three datasets: 1) LJSpeech\footnote{\url{https://keithito.com/LJ-Speech-Dataset/}} \cite{ljspeech17}, a single-speaker database which contains 13,100 English audio clips with a total of nearly 24 hours speech; 2) Biaobei\footnote{\url{https://www.data-baker.com/open_source.html}}, a Chinese speech corpus consists of 10,000 sentences (about 12 hours) from a Chinese speaker; 3) LibriTTS\footnote{\url{http://www.openslr.org/60}.} \cite{zen2019libritts}, an English dataset with 149,736 audio clips (about 245 hours) from 1,151 speakers (We only use \textit{train\_clean360} and \textit{train\_clean100}). For computational efficiency, we first use the syntactic graph builder to process the raw text of the whole dataset to construct syntactic graphs and record them in the disk. We then load the mini-batch along with the pre-constructed syntactic graph during training and testing. The raw text is transformed into a phoneme sequence using an open-sourced grapheme-to-phoneme tool. The ground truth mel-spectrograms are generated from the raw waveform with the frame size 1024 and the hop size 256. We compare SyntaSpeech against two state-of-the-art NAR-TTS models: \textit{PortaSpeech} and \textit{FastSpeech 2}. 
	
	\paragraph{Model Configuration} SyntaSpeech consists of a phoneme encoder, a linguistic encoder, two syntactic graph encoders (with the same structures), a variational generator, and a multi-length discriminator. The phoneme encoder and linguistic encoder are based on multiple feed-forward Transformer blocks, and the variational generator uses the same structure in PortaSpeech. The multi-length discriminator is a lightweight CNN that consists of multiple stacked convolutional layers with batch normalization and treats the input spectrogram as images. We put more detailed model configurations in Appendix B.1.
	
	\paragraph{Training and Evaluation} We train the SyntaSpeech on 1 Nvidia 2080Ti GPU with a batch size of 64 sentences. We use the Adam optimizer with $\beta_1=0.9$ ,$\beta_2=0.98$, $\epsilon=10^{-9}$ and follow the same learning rate schedule in \cite{vaswani2017attention}. It takes 320k steps for training until convergence. We use HiFi-GAN \cite{kong2020hifigan} as the vocoder in LJSpeech and Biaobei, and use Parallel WaveGAN \cite{yamamoto2020pwg} as the vocoder in LibriTTS. We conduct MOS (mean opinion score) and CMOS (comparative mean opinion score) evaluations on the test set via Amazon Mechanical Turk. We analyze the MOS and CMOS in two aspects: prosody (naturalness of pitch, energy, and duration) and audio quality (clarity, high-frequency and original timbre reconstruction), and score MOS-P/CMOS-P and MOS-Q/CMOS-Q corresponding to the MOS/CMOS of prosody and audio quality. We put more details about the subjective evaluation in Appendix B.2.
	
	\subsection{Performance}
	We compare the audio performance (MOS-P and MOS-Q) of our SyntaSpeech with other systems, including 1) \textit{GT}, the ground truth audio; 2)\textit{ GT (voc.)}, where we first convert the ground truth audio into mel-spectrograms, and then convert the mel-spectrograms back to audio using external vocoders; 3) \textit{FastSpeech2} \cite{ren2020fastspeech2}; 4) \textit{PortaSpeech} \cite{ren2021portaspeech}. We perform the experiments on three datasets as mentioned in Sec.\ref{sec:experimental_setup}. The results are shown in Table \ref{tab:mos-p} and \ref{tab:mos-q}. We observe that SyntaSpeech outperforms previous TTS models in both prosody (MOS-P) and audio quality (MOS-Q), which demonstrates its performance and robustness in multiple languages and multi-speaker 
	TTS tasks. As our SyntaSpeech follows the variational generator in PortaSpeech, we perform a case study to demonstrate that SyntaSpeech could generate more natural audio than its baseline PortaSpeech, using a variety of latent variables of VAE. The result is put in Appendix C.1. 
	
	\begin{table}
		\small
		\centering
		\begin{tabular}{lccc}
			\toprule
			Method  & LJSpeech & Biaobei & LibriTTS  \\
			\midrule
			\textit{GT}       & $4.32\pm0.09$  & $4.43\pm0.05$ & $4.32\pm 0.07$      \\
			\textit{GT (voc.)} & $4.26\pm0.09$  & $4.34\pm0.05$  & $4.29\pm0.07$   \\
			\midrule
			\textit{FastSpeech2}    & $3.85\pm0.12$  & $3.75\pm0.10$ & $3.98\pm 0.08$     \\
			\textit{PortaSpeech}   & $4.01\pm0.12$  & $3.90\pm0.10$  & $4.06\pm0.07$  \\
			\midrule
			\textit{SyntaSpeech} & \textbf{4.19} $\pm$ \textbf{0.10} & \textbf{4.12} $\pm$ \textbf{0.07}  & \textbf{4.18} $\pm$  \textbf{0.07}      \\
			\bottomrule
		\end{tabular}
		\caption{MOS-P evaluation on three datasets.}
		\label{tab:mos-p}
	\end{table}
	
	\begin{table}
		\small
		\centering
		\begin{tabular}{lccc}
			\toprule
			Method  & LJSpeech & Biaobei & LibriTTS  \\
			\midrule
			\textit{GT}       & $4.26\pm0.06$  & $4.46\pm0.05$ & $4.25\pm0.06$      \\
			\textit{GT (voc.)} & $4.17\pm0.08$  & $4.33\pm0.06$  &   $4.19\pm0.08$ \\
			\midrule
			\textit{FastSpeech2}    & $3.94\pm0.09$  & $3.82\pm0.09$ & $3.95\pm0.09$    \\
			\textit{PortaSpeech}   & $4.02\pm0.08$  & $4.05\pm0.08$  & $4.03\pm0.10$  \\
			\midrule
			\textit{SyntaSpeech} & \textbf{4.13} $\pm$ \textbf{0.08} & \textbf{4.19} $\pm$ \textbf{0.07} & \textbf{4.10} $\pm$ \textbf{0.08}     \\
			\bottomrule
		\end{tabular}
		\caption{MOS-Q evaluation on three datasets.}
		\label{tab:mos-q}
	\end{table}
	
\begin{figure*}[!t]
		\centering
		\subfloat[\textit{GT}]{\includegraphics[width=4.15cm]{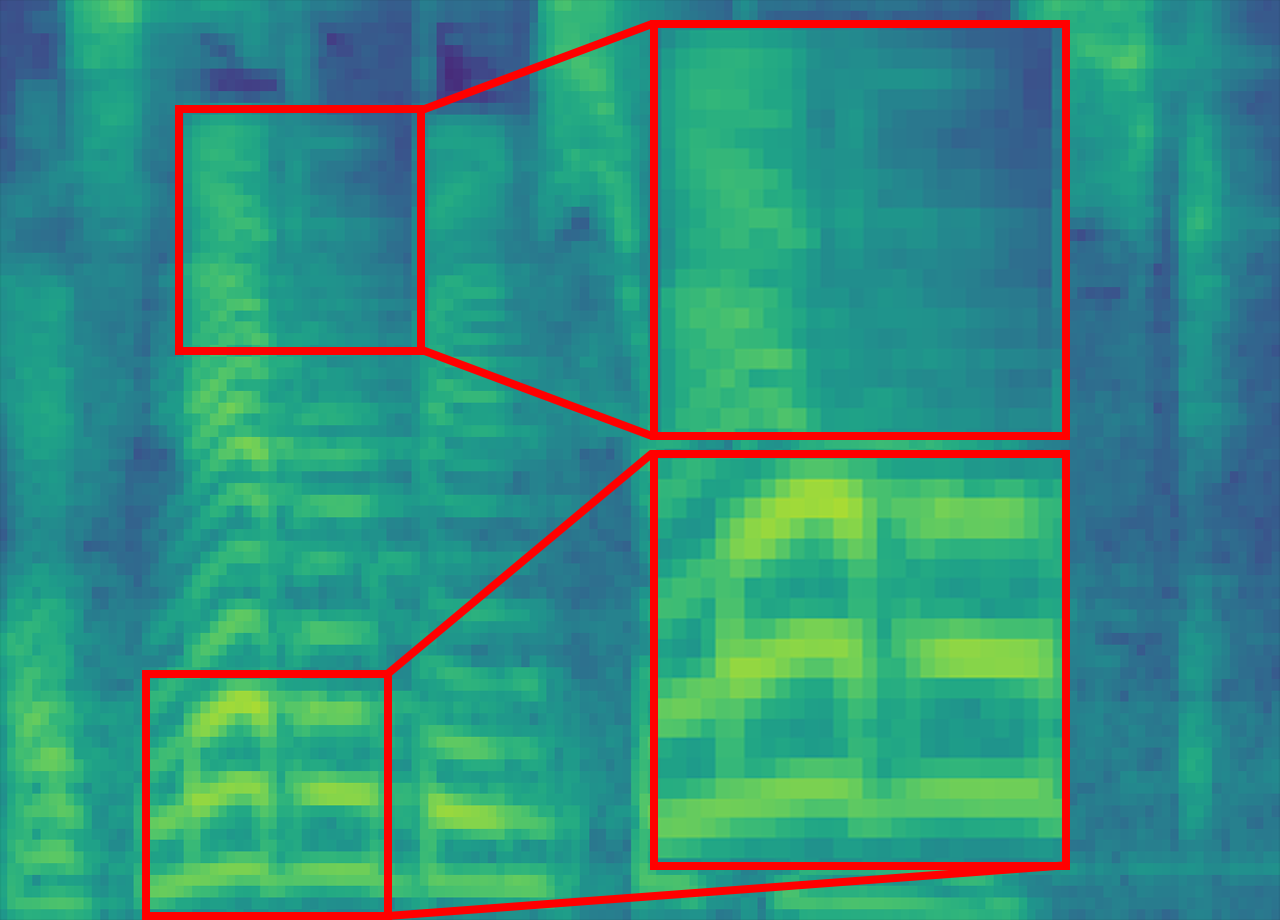}%
			\label{figure:mel_gt}}
		\hfil
		\subfloat[\textit{FastSpeech 2}]{\includegraphics[width=4.15cm]{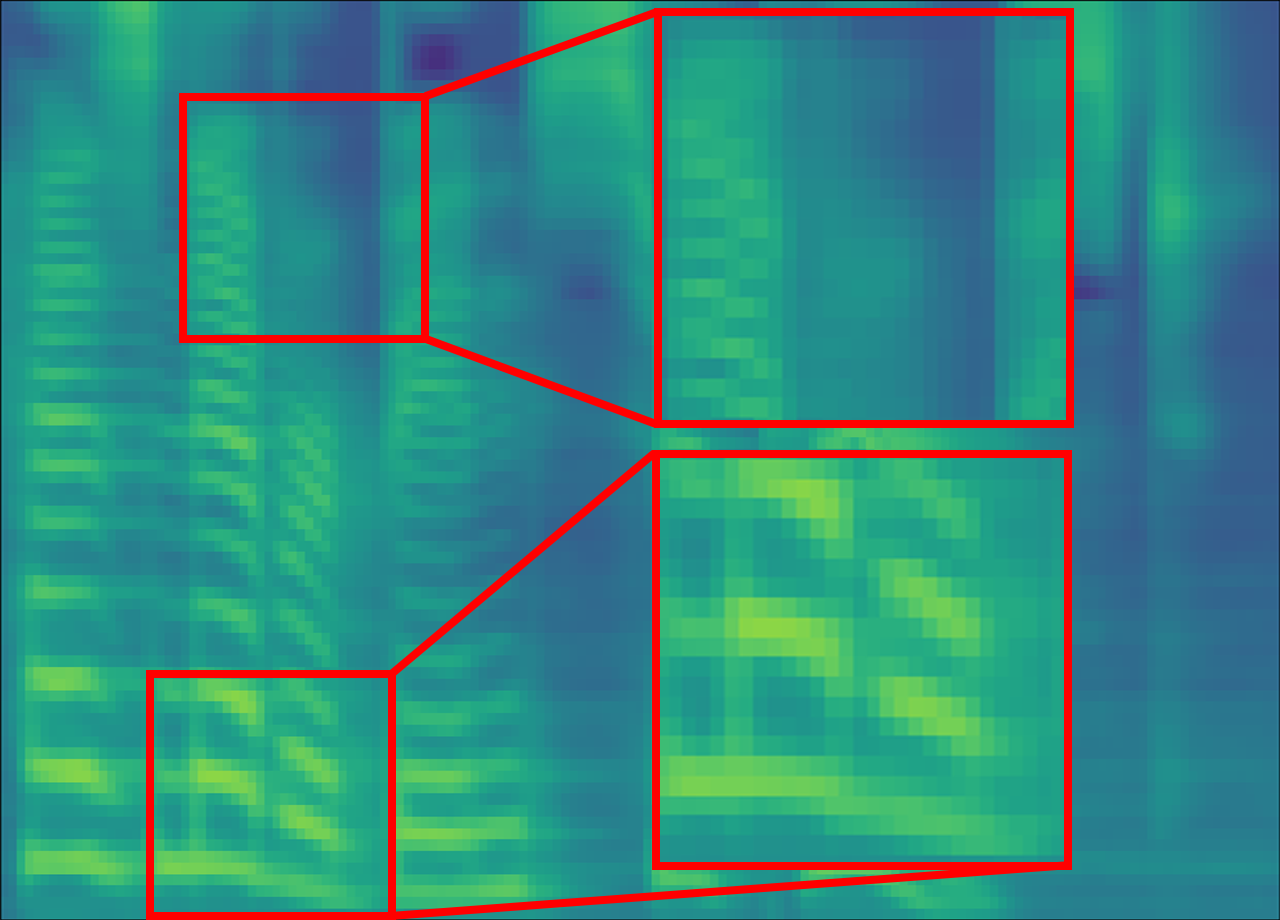}%
			\label{figure:mel_fs2}}
		\hfil
		\subfloat[\textit{PortaSpeech}]{\includegraphics[width=4.15cm]{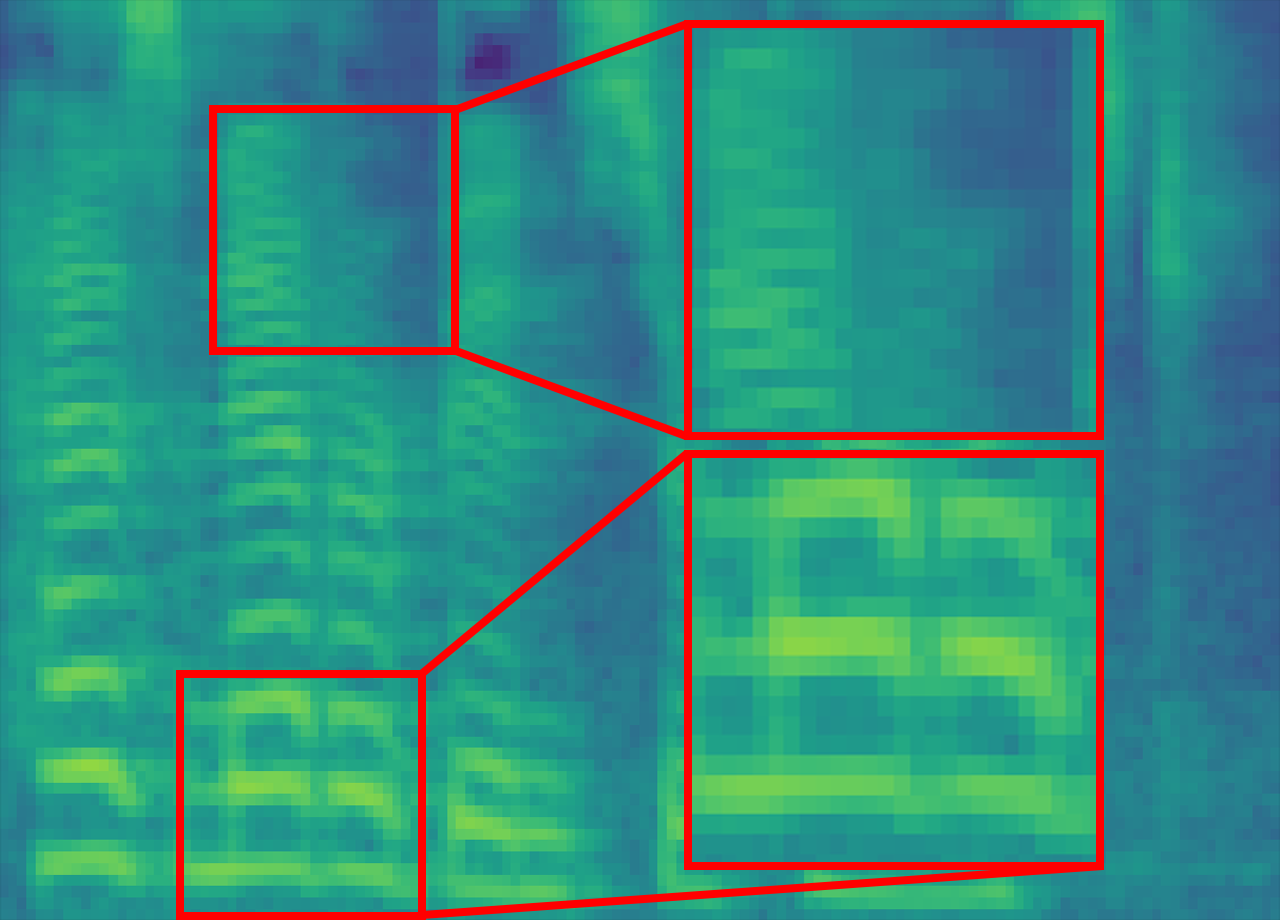}%
			\label{figure:mel_ps}}
		\hfil
		\subfloat[\textit{SyntaSpeech}]{\includegraphics[width=4.15cm]{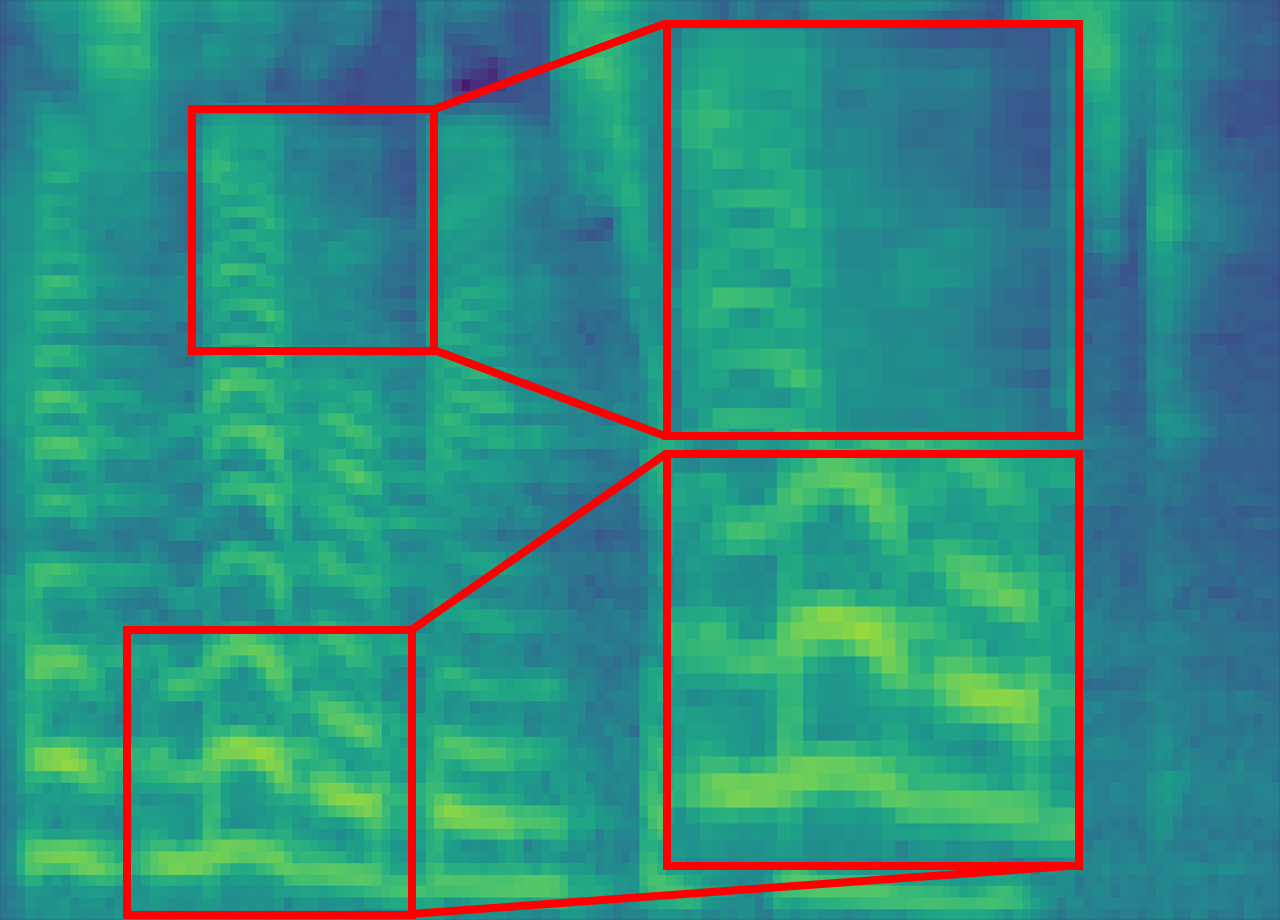}%
			\label{figure:mel_synta}}
		\caption{Visualizations of the mel-spectrograms generated by different TTS systems.	The corresponding text is \textit{"has never been surpassed"}.}
		\label{figure:mels}	
	\end{figure*}

	We then visualize the mel-spectrograms generated by the above systems in Fig.\ref{figure:mels}. We can see that SyntaSpeech can generate mel-spectrograms with realistic pitch contours (which result in expressive prosody) and rich details in frequency bins (which result in natural sounds). In conclusion, our experiments demonstrate that SyntaSpeech could synthesize expressive and high-quality audio. 
	
	\subsection{Ablation Studies}
	\subsubsection{Syntactic Graph Encoder}
\begin{figure}[!t]	
		\centering
		\includegraphics[width=8cm]{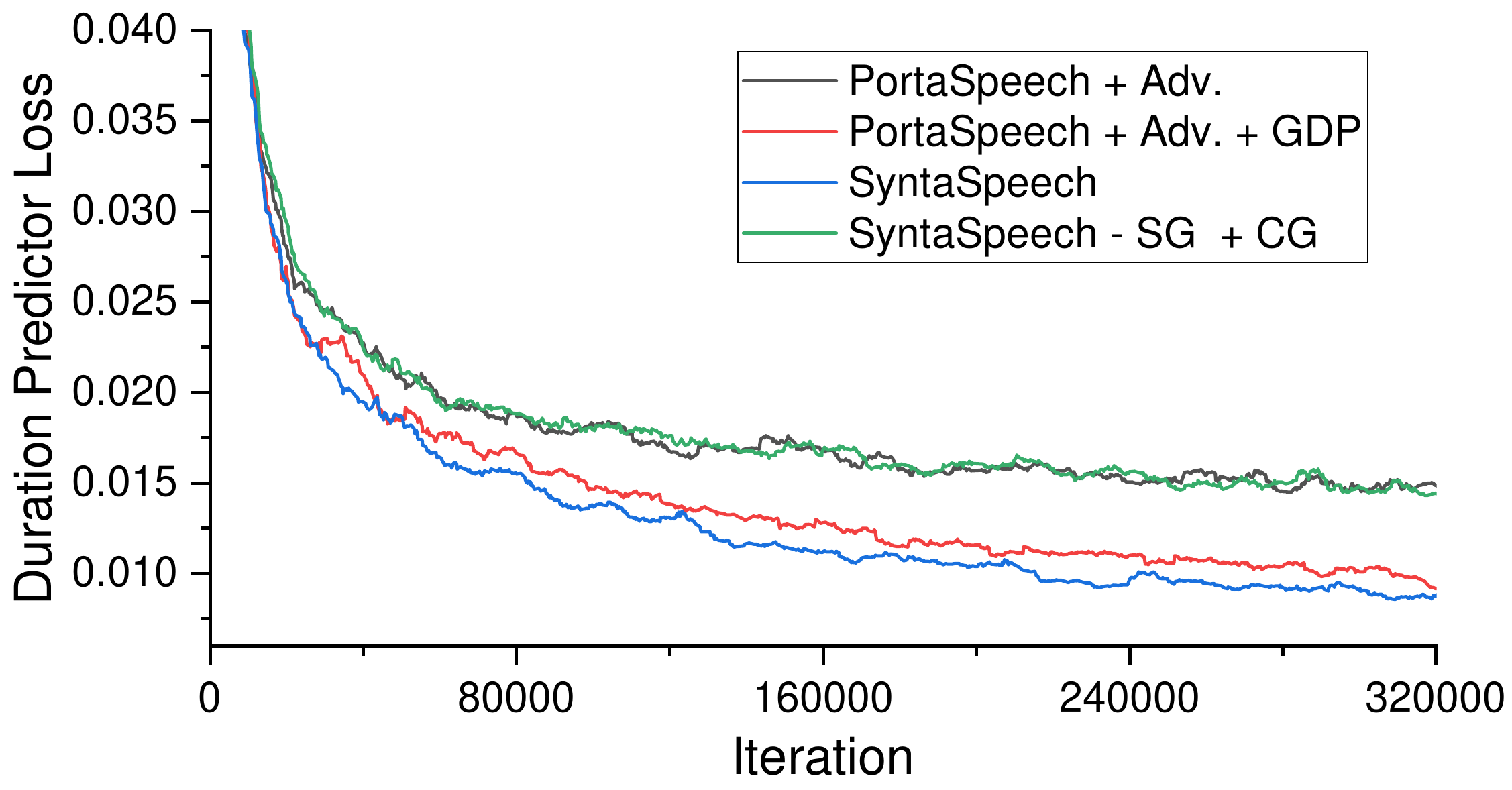}
		\caption{The duration predictor loss curves of several methods in LJSpeech. \textit{Adv} denotes multi-length adversarial training, \textit{GDP} denotes using graph encoder in duration predictor, \textit{CG} denotes using complete graph instead of the syntactic graph (\textit{SG}).}
		\label{figure:LJSpeech_DP_loss}	
	\end{figure}

	We first analyze the effectiveness of the syntactic graph encoder to improve prosody from the perspective of training objectives. The learning curves of duration predictor loss\footnote{The duration predictor loss is the mean squared error between the logarithmic predicted word-level duration and the ground truth.} in LJSpeech are shown in Fig.\ref{figure:LJSpeech_DP_loss}. We observe that introducing a graph encoder in the duration predictor (\textit{GDP}) could significantly improve the convergence. And SyntaSpeech, which is equivalent to \textit{(PortaSpeech + Adv. + GDP + GPF)}, where \textit{GPF} denotes using graph encoder in the prior flow, could further improve the performance. We also demonstrate that the improvement is brought by the syntactic information, as replacing the syntactic graph with the complete graph in SyntaSpeech leads to a similar curve to PortaSpeech. We put more objective evaluations in Appendix C.2.
	
	We then perform CMOS evaluation to demonstrate the effectiveness of \textit{syntactic graph encoder} in SyntaSpeech to improve prosody prediction. The results are shown in Table.\ref{tab:cmos-p}. We can see that CMOS-P drops when removing graph encoder in duration predictor (\textit{- GDP}) or prior flow (\textit{- GPF}), and replacing syntactic graph with complete graph (\textit{- SG + CG}) leads to the largest CMOS-P degradation. A similar experiment that tests CMOS-Q can be found in Appendix C.3, in which we find that syntactic graph encoder has fewer impacts on the audio quality.
	
	\begin{table}
		\small
		\centering
		\begin{tabular}{lccc}
			\toprule
			Settings  & LJSpeech & Biaobei & LibriTTS  \\
			\toprule
			\textit{SyntaSpeech}       & $0.000$  & $0.000$ & $0.000$      \\
			\toprule
			\textit{- GDP}    & $-0.131$  & $-0.092$ & $-0.119$     \\
			\textit{- GPF}   & $-0.069$  & $-0.118$  & $-0.059$  \\
			\textit{- GDP - GPF}   & $-0.152$  & $-0.142$  & $-0.168$  \\
			\textit{- SG + CG} & $-0.160$ & $-0.109$  &  $-0.188$  \\
			\bottomrule
		\end{tabular}
		\caption{CMOS-P comparisons for ablation studies.}
		\label{tab:cmos-p}
	\end{table}

	\subsubsection{Adversarial Training}
	To demonstrate the effectiveness of adversarial training, we perform a CMOS test on PortaSpeech/SyntaSpeech with the multi-length adversarial training and the post-net. As can be seen in Table.\ref{tab:cmos-q}, both in PortaSpeech and our SyntaSpeech, multi-length adversarial training achieves better audio quality (CMOS-Q) than the flow-based post-net. We also compare the COMS-P, as can be found in Appendix C.4, in which we find that adversarial training also has slight improvements on the audio prosody.
	
	\begin{table}
		\small
		\centering
		\begin{tabular}{lccc}
			\toprule
			Settings  & LJSpeech & Biaobei & LibriTTS  \\
			\toprule
			\textit{PortaSpeech}       & $0.000$  & $0.000$ & $0.000$      \\
			\textit{- PN + Adv.}    & $0.071$  & $0.088$ & $0.050$     \\
			\midrule
			\textit{SyntaSpeech}   & $0.000$  & $0.000$  & $0.000$  \\
			\textit{- Adv. + PN}   & $-0.060$  & $0.166$  & $-0.039$  \\
			\bottomrule
		\end{tabular}
		\caption{CMOS-Q comparisons for ablation studies. \textit{PN} denotes post-net in PortaSpeech, and \textit{Adv} means our adversarial training.}
		\label{tab:cmos-q}
	\end{table}
	
	\section{Conclusion}
	\label{sec:conclusion}
	In this paper, we proposed SyntaSpeech, a syntax-aware and generative adversarial text-to-speech model. SyntaSpeech builds the syntactic graph from the dependency tree of the raw text, then extracts valuable syntactic information with graph convolution on the syntactic graph to improve the prosody prediction in the NAR-TTS model. We also introduced multi-length adversarial training to improve the audio quality and simplify the model architecture. We have demonstrated the performance and generalization ability of SyntaSpeech on three datasets (English, Chinese, and multi-speaker, respectively) and conducted comprehensive ablation studies to verify the effectiveness of each component in our model. For future work, we will explore the potential of syntax-aware models in other tasks, such as voice conversion and singing voice generation.
	
	\newpage
	\bibliographystyle{named}
	\bibliography{ijcai22}

\begin{thebibliography}{}

\bibitem[\protect\citeauthoryear{Bińkowski \bgroup \em et al.\egroup
  }{2020}]{binkowski2019gan-tts}
Mikołaj Bińkowski, Jeff Donahue, Sander Dieleman, Aidan Clark, Erich Elsen,
  Norman Casagrande, Luis~C. Cobo, and Karen Simonyan.
\newblock High fidelity speech synthesis with adversarial networks.
\newblock In {\em ICLR}, 2020.

\bibitem[\protect\citeauthoryear{Chen \bgroup \em et al.\egroup
  }{2020}]{chen2020hifisinger}
Jiawei Chen, Xu~Tan, Jian Luan, Tao Qin, and Tie-Yan Liu.
\newblock Hifisinger: Towards high-fidelity neural singing voice synthesis.
\newblock {\em arXiv preprint arXiv:2009.01776}, 2020.

\bibitem[\protect\citeauthoryear{Cho \bgroup \em et al.\egroup
  }{2014}]{cho2014gru}
Kyunghyun Cho, Bart van Merrienboer, Çaglar G{\"u}lçehre, Dzmitry Bahdanau,
  Fethi Bougares, Holger Schwenk, and Yoshua Bengio.
\newblock Learning phrase representations using rnn encoder–decoder for
  statistical machine translation.
\newblock In {\em EMNLP}, 2014.

\bibitem[\protect\citeauthoryear{Devlin \bgroup \em et al.\egroup
  }{2019}]{devlin2019bert}
Jacob Devlin, Ming{-}Wei Chang, Kenton Lee, and Kristina Toutanova.
\newblock Bert: Pre-training of deep bidirectional transformers for language
  understanding.
\newblock In {\em NAACL-HLT}, 2019.

\bibitem[\protect\citeauthoryear{Donahue \bgroup \em et al.\egroup
  }{2021}]{donahue2020eats}
Jeff Donahue, Sander Dieleman, Mikolaj Binkowski, Erich Elsen, and Karen
  Simonyan.
\newblock End-to-end adversarial text-to-speech.
\newblock In {\em ICLR}, 2021.

\bibitem[\protect\citeauthoryear{Hirschberg and
  Rambow}{2001}]{hirschberg2001prosody-and-syntax}
Julia Hirschberg and Owen Rambow.
\newblock Learning prosodic features using a tree representation.
\newblock In {\em ECSCT}, 2001.

\bibitem[\protect\citeauthoryear{Ito and Johnson}{2017}]{ljspeech17}
Keith Ito and Linda Johnson.
\newblock The lj speech dataset.
\newblock \url{https://keithito.com/LJ-Speech-Dataset/}, 2017.

\bibitem[\protect\citeauthoryear{Kim \bgroup \em et al.\egroup
  }{2020}]{kim2020glowtts}
Jaehyeon Kim, Sungwon Kim, Jungil Kong, and Sungroh Yoon.
\newblock Glow-tts: A generative flow for text-to-speech via monotonic
  alignment search.
\newblock In {\em NIPS}, 2020.

\bibitem[\protect\citeauthoryear{Kim \bgroup \em et al.\egroup
  }{2021}]{kim2021vits}
Jaehyeon Kim, Jungil Kong, and Juhee Son.
\newblock Conditional variational autoencoder with adversarial learning for
  end-to-end text-to-speech.
\newblock In {\em ICML}, 2021.

\bibitem[\protect\citeauthoryear{Kong \bgroup \em et al.\egroup
  }{2020}]{kong2020hifigan}
Jungil Kong, Jaehyeon Kim, and Jaekyoung Bae.
\newblock Hifi-gan: Generative adversarial networks for efficient and high
  fidelity speech synthesis.
\newblock In {\em NIPS}, 2020.

\bibitem[\protect\citeauthoryear{Li \bgroup \em et al.\egroup
  }{2016}]{li2015gated}
Yujia Li, Richard Zemel, Marc Brockschmidt, and Daniel Tarlow.
\newblock Gated graph sequence neural networks.
\newblock In {\em ICLR}, 2016.

\bibitem[\protect\citeauthoryear{Liu \bgroup \em et al.\egroup
  }{2021}]{liu2021graphspeech}
Rui Liu, Berrak Sisman, and Haizhou Li.
\newblock Graphspeech: Syntax-aware graph attention network for neural speech
  synthesis.
\newblock In {\em ICASSP}, 2021.

\bibitem[\protect\citeauthoryear{Mishra \bgroup \em et al.\egroup
  }{2015}]{mishra2015intonational}
Taniya Mishra, Yeon-jun Kim, and Srinivas Bangalore.
\newblock Intonational phrase break prediction for text-to-speech synthesis
  using dependency relations.
\newblock In {\em ICASSP}, 2015.

\bibitem[\protect\citeauthoryear{Peng \bgroup \em et al.\egroup
  }{2020}]{peng2020non}
Kainan Peng, Wei Ping, Zhao Song, and Kexin Zhao.
\newblock Non-autoregressive neural text-to-speech.
\newblock In {\em ICML}, 2020.

\bibitem[\protect\citeauthoryear{Ping \bgroup \em et al.\egroup
  }{2018}]{ping2018deepvoice3}
Wei Ping, Kainan Peng, Andrew Gibiansky, Sercan~O Arik, Ajay Kannan, Sharan
  Narang, Jonathan Raiman, and John Miller.
\newblock Deep voice 3: Scaling text-to-speech with convolutional sequence
  learning.
\newblock In {\em ICLR}, 2018.

\bibitem[\protect\citeauthoryear{Qi \bgroup \em et al.\egroup
  }{2020}]{qi2020stanza}
Peng Qi, Yuhao Zhang, Yuhui Zhang, Jason Bolton, and Christopher~D. Manning.
\newblock Stanza: A {Python} natural language processing toolkit for many human
  languages.
\newblock In {\em ACL}, 2020.

\bibitem[\protect\citeauthoryear{Ren \bgroup \em et al.\egroup
  }{2019}]{ren2019fastspeech}
Yi~Ren, Yangjun Ruan, Xu~Tan, Tao Qin, Sheng Zhao, Zhou Zhao, and Tie-Yan Liu.
\newblock Fastspeech: fast, robust and controllable text to speech.
\newblock In {\em NIPS}, pages 3171--3180, 2019.

\bibitem[\protect\citeauthoryear{Ren \bgroup \em et al.\egroup
  }{2021a}]{ren2020fastspeech2}
Yi~Ren, Chenxu Hu, Xu~Tan, Tao Qin, Sheng Zhao, Zhou Zhao, and Tie-Yan Liu.
\newblock Fastspeech 2: Fast and high-quality end-to-end text to speech.
\newblock In {\em ICLR}, 2021.

\bibitem[\protect\citeauthoryear{Ren \bgroup \em et al.\egroup
  }{2021b}]{ren2021portaspeech}
Yi~Ren, Jinglin Liu, and Zhou Zhao.
\newblock Portaspeech: Portable and high-quality generative text-to-speech.
\newblock In {\em NIPS}, 2021.

\bibitem[\protect\citeauthoryear{Sun \bgroup \em et al.\egroup
  }{2020}]{sun2020graphtts}
Aolan Sun, Jianzong Wang, Ning Cheng, Huayi Peng, Zhen Zeng, and Jing Xiao.
\newblock Graphtts: graph-to-sequence modelling in neural text-to-speech.
\newblock In {\em ICASSP}, 2020.

\bibitem[\protect\citeauthoryear{Sun \bgroup \em et al.\egroup
  }{2021}]{sun2021graphpb}
Aolan Sun, Jianzong Wang, Ning Cheng, Huayi Peng, Zhen Zeng, Lingwei Kong, and
  Jing Xiao.
\newblock Graphpb: Graphical representations of prosody boundary in speech
  synthesis.
\newblock In {\em SLT}, 2021.

\bibitem[\protect\citeauthoryear{{van den Oord} \bgroup \em et al.\egroup
  }{2016}]{oord2016wavenet}
Aäron {van den Oord}, Sander Dieleman, Heiga Zen, Karen Simonyan, Oriol
  Vinyals, Alex Graves, Nal Kalchbrenner, Andrew Senior, and Koray Kavukcuoglu.
\newblock {WaveNet: A Generative Model for Raw Audio}.
\newblock In {\em SSW}, 2016.

\bibitem[\protect\citeauthoryear{Vaswani \bgroup \em et al.\egroup
  }{2017}]{vaswani2017attention}
Ashish Vaswani, Noam Shazeer, Niki Parmar, Jakob Uszkoreit, Llion Jones,
  Aidan~N Gomez, {\L}ukasz Kaiser, and Illia Polosukhin.
\newblock Attention is all you need.
\newblock In {\em NIPS}, 2017.

\bibitem[\protect\citeauthoryear{Wang \bgroup \em et al.\egroup
  }{2017}]{wang2017tacotron}
Yuxuan Wang, R.~J. Skerry-Ryan, Daisy Stanton, Yonghui Wu, Ron~J. Weiss,
  Navdeep Jaitly, Zongheng Yang, Ying Xiao, Z.~Chen, Samy Bengio, Quoc~V. Le,
  Yannis Agiomyrgiannakis, Robert A.~J. Clark, and Rif~A. Saurous.
\newblock Tacotron: Towards end-to-end speech synthesis.
\newblock In {\em INTERSPEECH}, 2017.

\bibitem[\protect\citeauthoryear{Yamamoto \bgroup \em et al.\egroup
  }{2020}]{yamamoto2020pwg}
Ryuichi Yamamoto, Eunwoo Song, and Jae-Min Kim.
\newblock Parallel wavegan: A fast waveform generation model based on
  generative adversarial networks with multi-resolution spectrogram.
\newblock In {\em ICASSP}, 2020.

\bibitem[\protect\citeauthoryear{Zen \bgroup \em et al.\egroup
  }{2019}]{zen2019libritts}
Heiga Zen, Viet Dang, Rob Clark, Yu~Zhang, Ron~J Weiss, Ye~Jia, Zhifeng Chen,
  and Yonghui Wu.
\newblock Libritts: A corpus derived from librispeech for text-to-speech.
\newblock {\em arXiv preprint arXiv:1904.02882}, 2019.

\bibitem[\protect\citeauthoryear{Zhou \bgroup \em et al.\egroup
  }{2021}]{zhou2021rgnn}
Yixuan Zhou, Changhe Song, Jingbei Li, Zhiyong Wu, and Helen Meng.
\newblock Dependency parsing based semantic representation learning with graph
  neural network for enhancing expressiveness of text-to-speech.
\newblock {\em arXiv preprint arXiv:2104.06835}, 2021.

\end{thebibliography}
	\newpage
	\appendix
	\section{Details of models}
	In this section, we describe details in the syntactic graph encoder and multi-length discriminator. For details in linguistic encoder and variational generator, one could refer to PortaSpeech \cite{ren2021portaspeech}. We also describe our modifications on PortaSpeech to support the multi-speaker task.
	
	\subsection{Syntactic Graph Encoder}
	
	\paragraph{External Dependency Parser}
	We use Stanza \cite{qi2020stanza} to extract dependency trees of English and Chinese text, then process the dependency tree into the syntactic graph following the method described in Sec.3.1. Stanza also supports other human languages such as French, Spanish, and Japanese. Note that we preprocess the Chinese text by separating the Chinese characters of each two adjacent words with a blank separator. We find it helps to improve the performance of Stanza's dependency parser for Chinese datasets.
	
	\paragraph{Batched Graph Convolution}
	For computational efficiency, we first use the syntactic graph builder to process the raw text of the whole dataset to construct syntactic graphs and record them in the disk. We then load the mini-batch along with the pre-constructed syntactic graph during training and testing. To execute graph convolution on the syntactic graphs in parallel, we integrate the syntactic graphs into one big graph before the graph convolution. After the graph convolution, the node embedding of the integrated graph is split into the original batch format and fed into the TTS model.
	
	\paragraph{Stop Gradient to Backbone}
	To improve the training stability, we disable the gradient backpropagation from the syntactic graph encoder to the phoneme encoder, so that the loss of duration predictor and prior flow do not directly affect that of the phoneme encoder. As shown in Fig.\ref{figure:disable_grad}, enabling the gradient causes the variational generator to converge too fast, and greatly slows down the training of mel-reconstruction (L1) loss. This may be because when the gradient of the variational generator can be directly propagated to the upstream of the network through the syntactic graph encoder, it may distract the phoneme encoder from its original training goal.
	
	\begin{figure}[!t]
		\centering
		\subfloat[KL Loss of variational generator]{\includegraphics[scale=0.4]{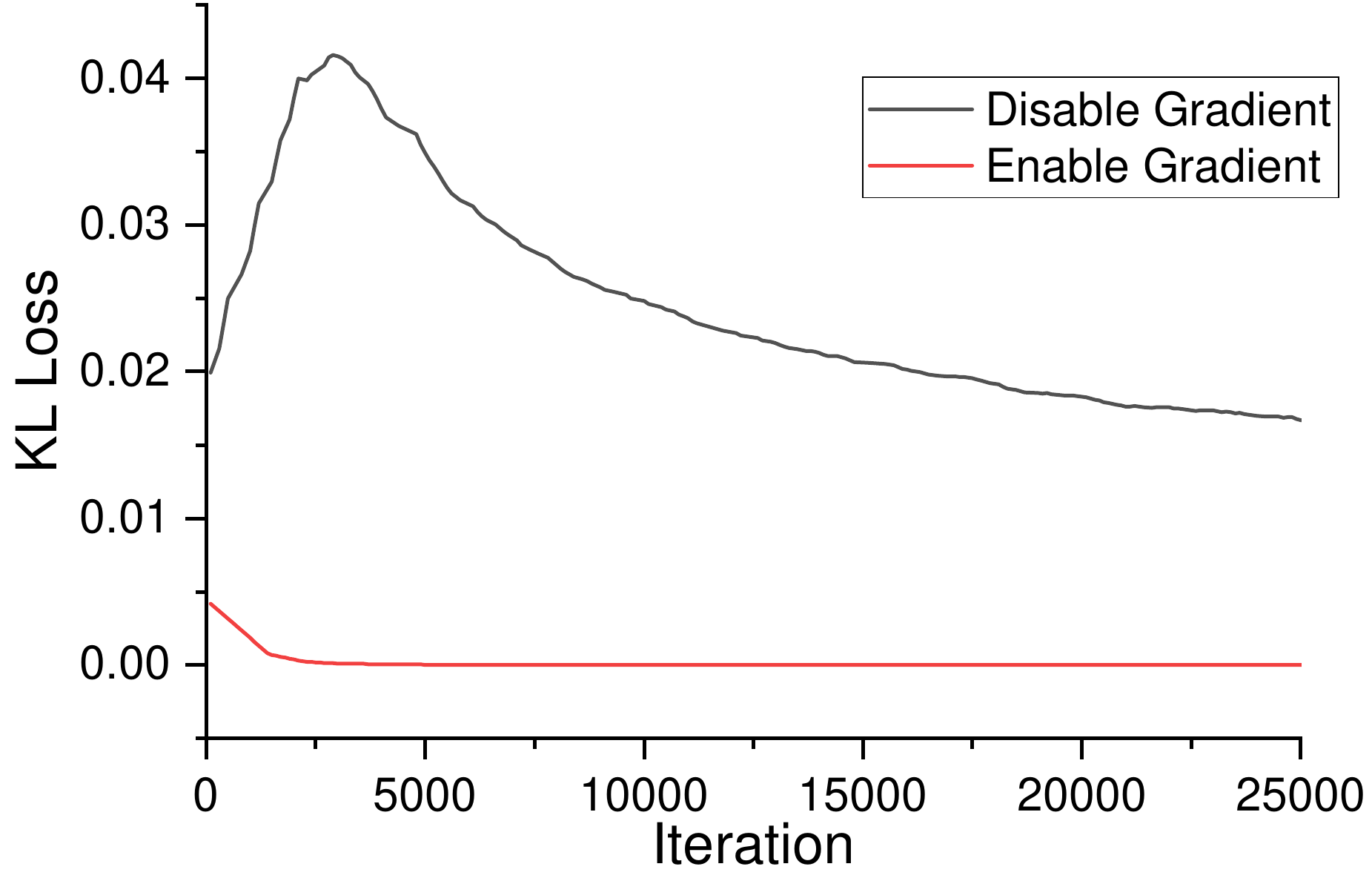}%
			\label{figure:ml_disc1}}
		\hfil
		\subfloat[L1 Loss]{\includegraphics[scale=0.4]{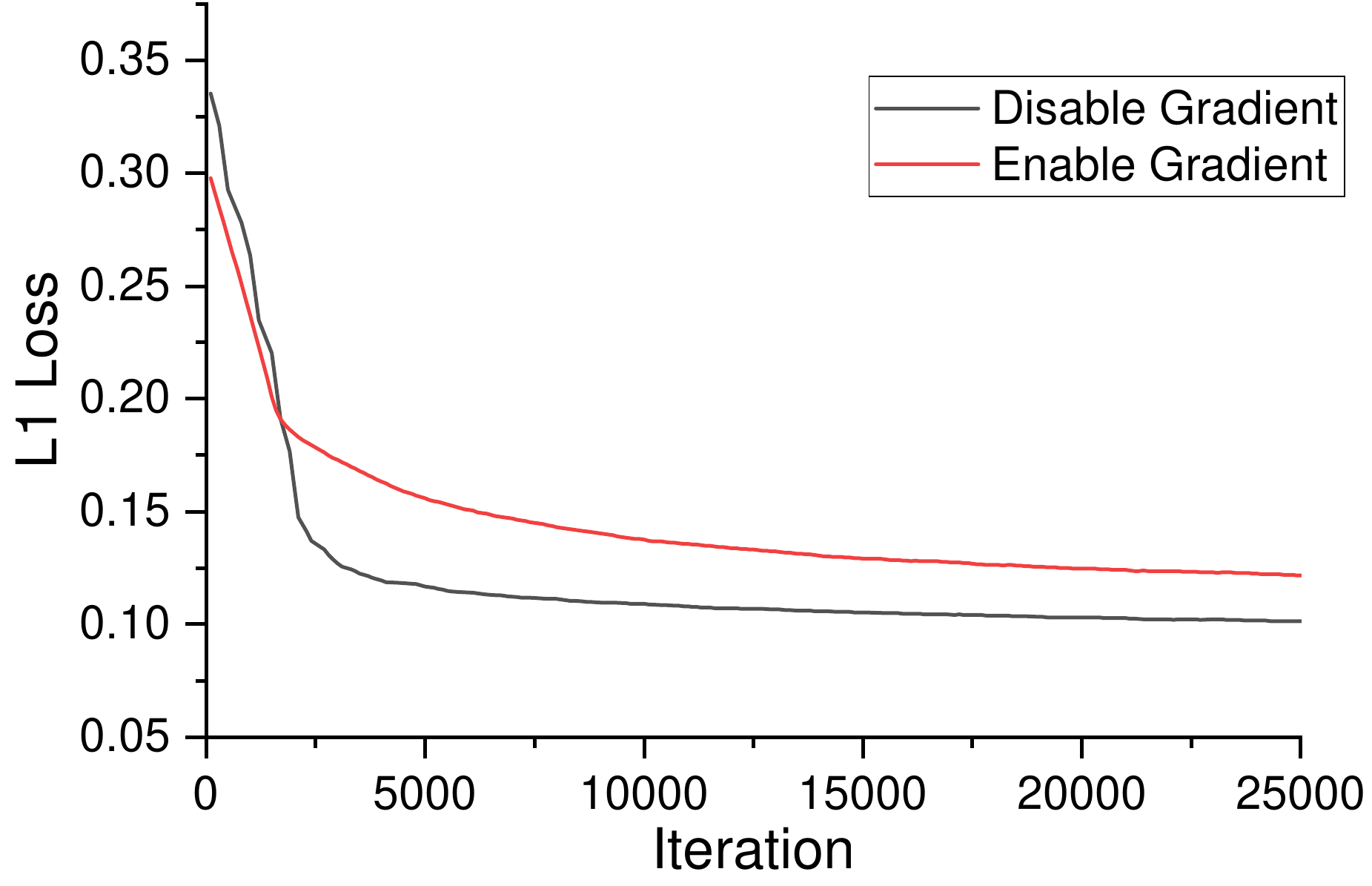}%
			\label{figure:ml_disc2}}
		\caption{The learning curves of SyntaSpeech that enables or disables the gradient from syntactic graph encoder.}
		\label{figure:disable_grad}	
	\end{figure}
	
	\subsection{Multi-Length Discriminator}
	The multi-length discriminator is an ensemble of multiple CNN-based discriminators which evaluates the mel-spectrogram based on random windows of different lengths, which is shown in Fig.\ref{figure:ml_disc1}. In our experiments, we train three CNN-based discriminators which observe random mel-spectrogram clips of length 32, 64, and 128 frames. The structure of the CNN-based discriminator is shown in Fig.\ref{figure:ml_disc2}. It consists of $N+1$ 2D-convolutional
	layers, each of which is followed by Leaky ReLU activation and DropOut. The latter $N$ convolutional layers are additionally followed by an instance norm layer. After the convolutional layers, a linear layer projects the hidden states of the mel-spectrogram slice to a scalar, which is the prediction that the input mel-spectrogram is true or fake. In our experiments, we set $N=2$.
	
	\begin{figure*}[!t]
		\centering
		\subfloat[Multi-Length Discriminator]{\includegraphics[scale=0.7]{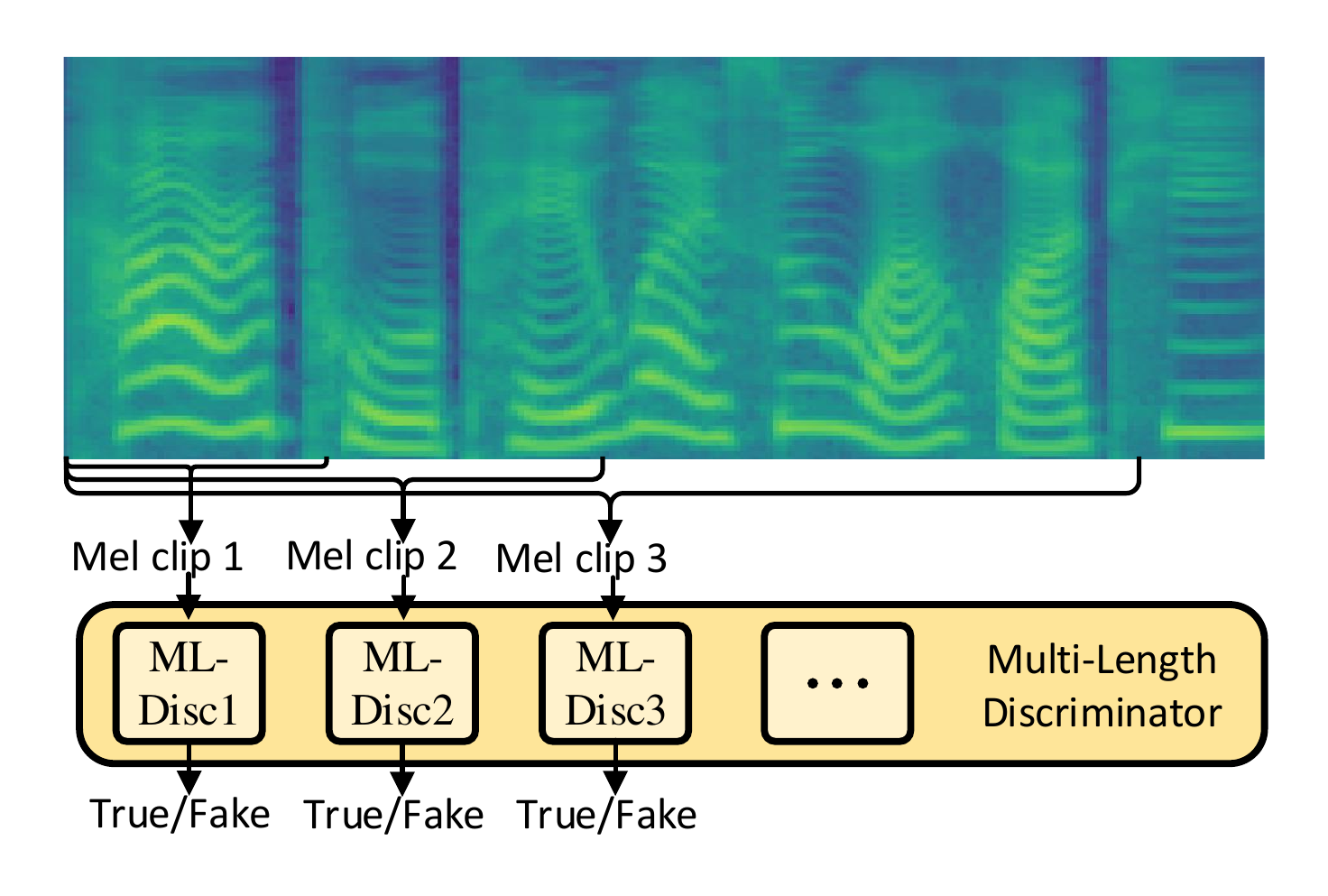}%
			\label{figure:ml_disc1}}
		\hfil
		\subfloat[CNN-Based Discriminator]{\includegraphics[scale=0.44]{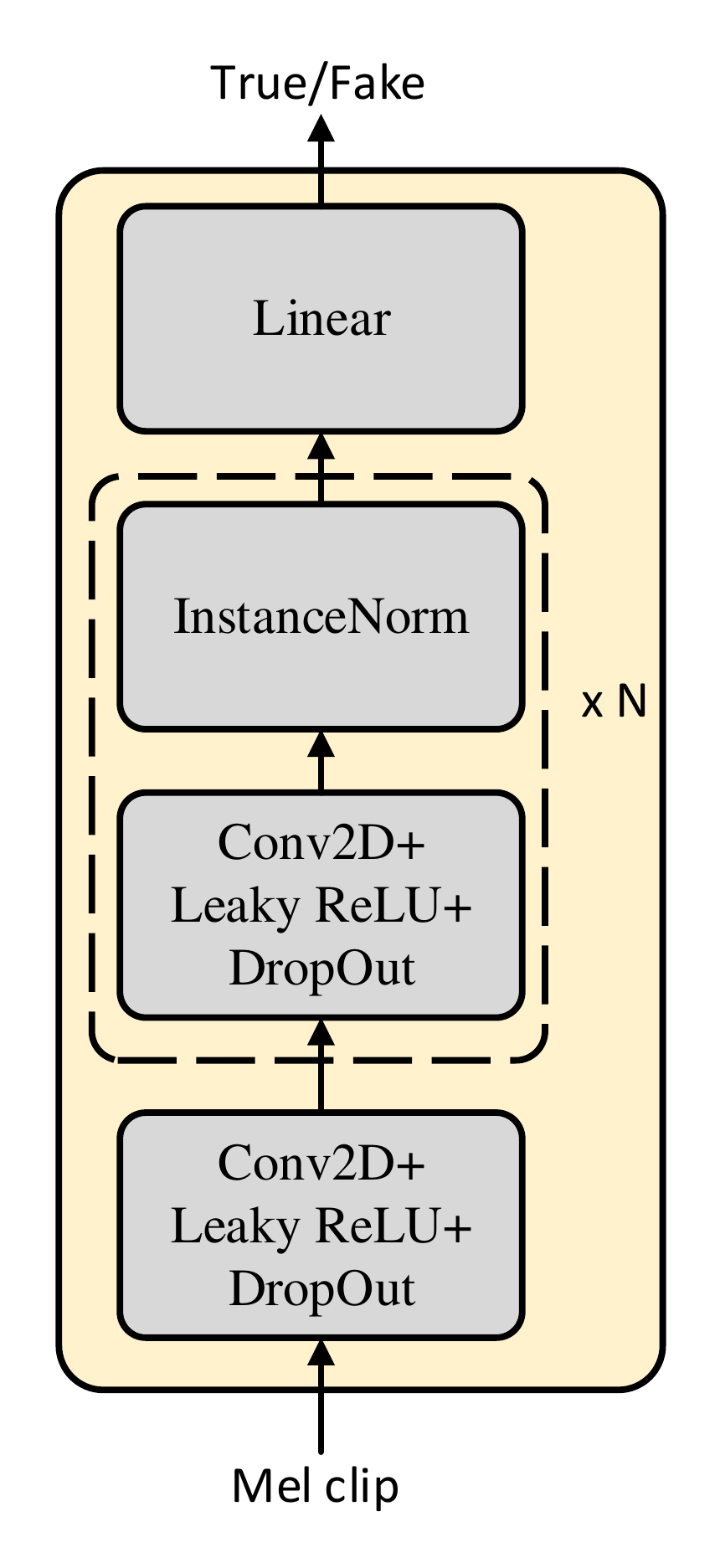}%
			\label{figure:ml_disc2}}
		\caption{The overall structure for multi-length discriminator.}
		\label{figure:network_structure}	
	\end{figure*}

	\subsection{Settings for Multi-Speaker Tasks}
	As vanilla PortaSpeech is tested on LJSpeech, which is a single-speaker English dataset. We make small modifications to support the multi-speaker task. To be specific, we additionally learn a group of speaker embedding to represent the speakers' timbre, pronunciation habits, and other features. The speaker embedding is fed into the input of variational generator and duration predictor and can be jointly optimized with backpropagation.

	\section{Detailed Experimental Settings}
	
	\subsection{Model Configurations}
	We list the hyper-parameters of SyntaSpeech in Tab. \ref{tab:model_configs}.
	
	\begin{table*}[htbp]
		\centering
		\caption{Add caption}
		\begin{tabular}{ccc|c}
			\hline
			\multicolumn{2}{c|}{Hyper-parameter} & SyntaSpeech & Number of parameters \bigstrut\\
			\hline
			\multicolumn{1}{c|}{\multirow{5}[2]{*}{Phoneme Encoder/ Linguistic Encoder}} & \multicolumn{1}{l|}{Phoneme embedding hidden size} & 192   & \multirow{5}[2]{*}{7.457M} \bigstrut[t]\\
			\multicolumn{1}{c|}{} & \multicolumn{1}{l|}{Word/phoneme encoder layers} & 4     &  \\
			\multicolumn{1}{c|}{} & \multicolumn{1}{l|}{Hidden size} & 192   &  \\
			\multicolumn{1}{c|}{} & \multicolumn{1}{l|}{Conv1D kernel} & 5     &  \\
			\multicolumn{1}{c|}{} & \multicolumn{1}{l|}{Conv1D filter size} & 768   &  \bigstrut[b]\\
			\hline
			\multicolumn{1}{c|}{\multirow{2}[2]{*}{Speaker Embedding (For LibriTTS)}} & \multicolumn{1}{l|}{Number of speakers} & 2320  & \multirow{2}[2]{*}{0.445M} \bigstrut[t]\\
			\multicolumn{1}{c|}{} & \multicolumn{1}{l|}{Hidden size} & 192   &  \bigstrut[b]\\
			\hline
			\multicolumn{1}{c|}{\multirow{3}[2]{*}{Syntactic Graph Encoder}} & \multicolumn{1}{l|}{Graph Convolutional Layers} & 2     & \multirow{3}[2]{*}{1.778M (0.889M*2)} \bigstrut[t]\\
			\multicolumn{1}{c|}{} & \multicolumn{1}{l|}{Hidden size} & 192   &  \\
			\multicolumn{1}{c|}{} & \multicolumn{1}{l|}{Number of encoders} & 2     &  \bigstrut[b]\\
			\hline
			\multicolumn{1}{c|}{\multirow{8}[2]{*}{Variational Generator}} & \multicolumn{1}{l|}{Encoder Layers} & 8     & \multirow{8}[2]{*}{7.516M} \bigstrut[t]\\
			\multicolumn{1}{c|}{} & \multicolumn{1}{l|}{Decoder Layers} & 4     &  \\
			\multicolumn{1}{c|}{} & \multicolumn{1}{l|}{Encoder/Decoder Conv1D Kernel} & 5     &  \\
			\multicolumn{1}{c|}{} & \multicolumn{1}{l|}{Encoder/Decoder Conv1D channel size} & 192   &  \\
			\multicolumn{1}{c|}{} & \multicolumn{1}{l|}{Latent Size } & 16    &  \\
			\multicolumn{1}{c|}{} & \multicolumn{1}{l|}{Prior Flow Layers} & 4     &  \\
			\multicolumn{1}{c|}{} & \multicolumn{1}{l|}{Prior Flow Conv1D Kernel} & 3     &  \\
			\multicolumn{1}{c|}{} & \multicolumn{1}{l|}{Prior Flow Conv1D Channel Size} & 64    &  \bigstrut[b]\\
			\hline
			\multicolumn{1}{c|}{\multirow{4}[2]{*}{Multi-Length Discriminator}} & \multicolumn{1}{l|}{Number of CNN-based Discriminators} & 3     & \multirow{4}[2]{*}{0.927M} \bigstrut[t]\\
			\multicolumn{1}{c|}{} & \multicolumn{1}{l|}{window size} & 32,64,128 &  \\
			\multicolumn{1}{c|}{} & \multicolumn{1}{l|}{Conv2D layers} & 3     &  \\
			\multicolumn{1}{c|}{} & \multicolumn{1}{l|}{Hidden size} & 192   &  \bigstrut[b]\\
			\hline
			\multicolumn{3}{c|}{Total Number of Parameters} & 18.123M \bigstrut\\
			\hline
		\end{tabular}%
		\label{tab:model_configs}%
	\end{table*}%

	\subsection{Details in Subjective Evaluation}
\begin{figure*}[!t]
		\centering
		\subfloat[Screenshot of MOS-P testing]{\includegraphics[scale=0.25]{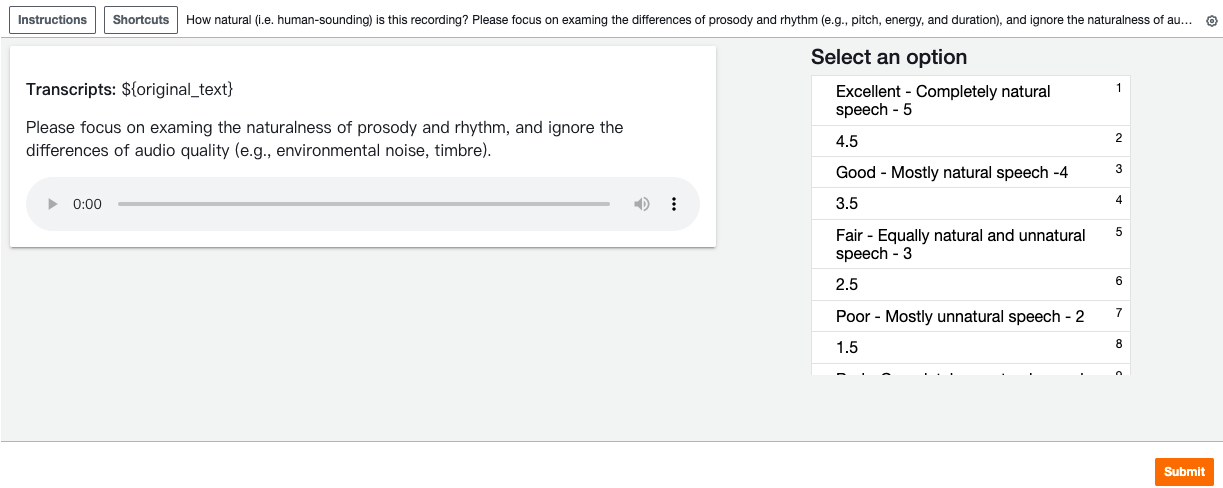}%
			\label{figure:screenshot_mos-p}}
		\hfil
		\subfloat[Screenshot of MOS-Q testing]{\includegraphics[scale=0.25]{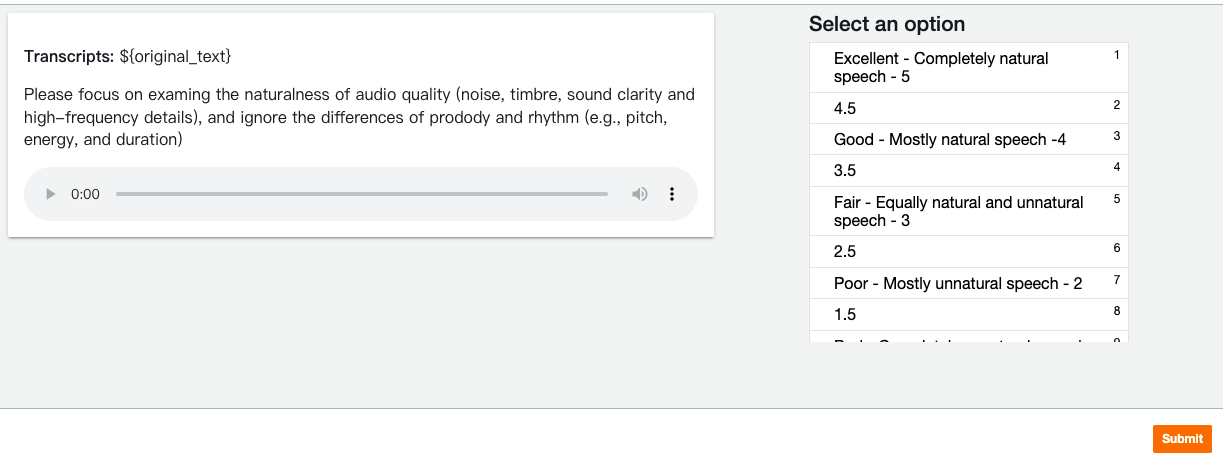}%
			\label{figure:screenshot_mos-q}}
		\hfil
		\subfloat[Screenshot of CMOS-P testing]{\includegraphics[scale=0.25]{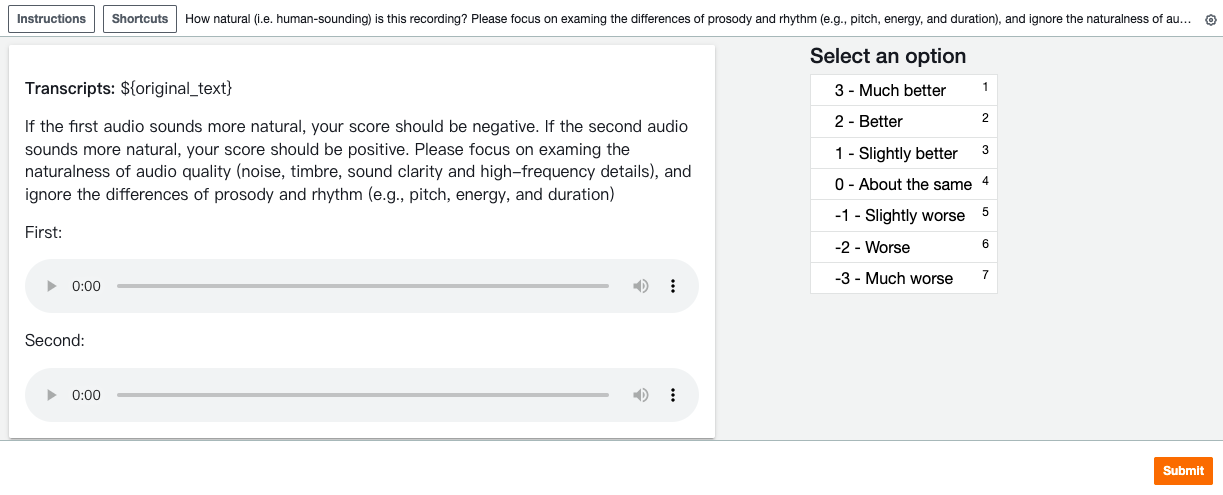}%
			\label{figure:screenshot_cmos-p}}
		\hfil
		\subfloat[Screenshot of CMOS-Q testing]{\includegraphics[scale=0.25]{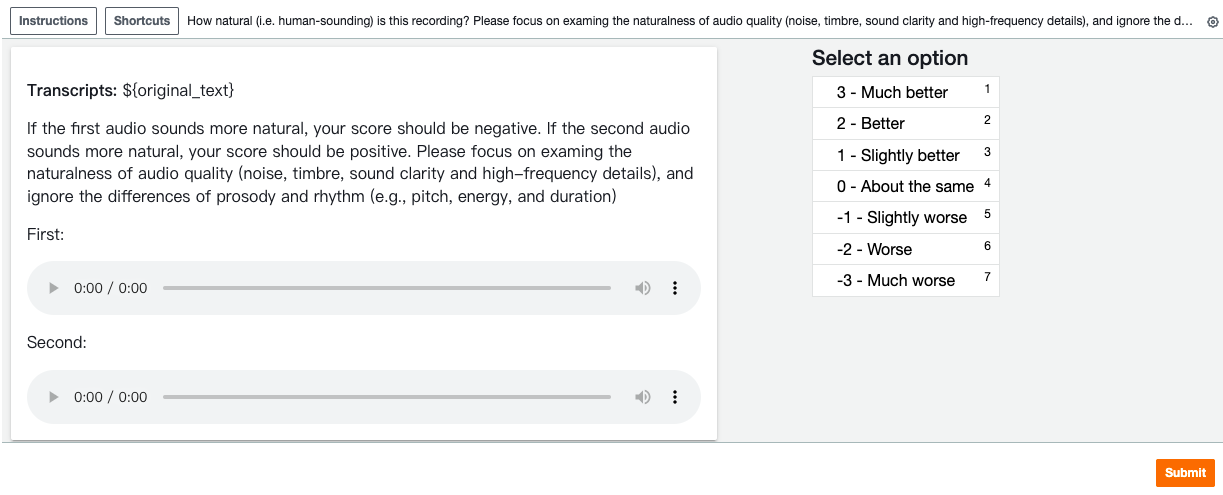}%
			\label{figure:screenshot_cmos-q}}
		\caption{Screenshots of our subjective evaluations on MTurk.}
		\label{figure:screenshots}	
	\end{figure*}

	We perform the subjective evaluation on Amazon Mechanical Turk (MTurk). For each tested dataset, we randomly select 20 texts from the test set and use the TTS systems to generate the audio samples. Each audio has been listened to by at least 6 listeners. For MOS, each tester is asked to evaluate the subjective naturalness of a sentence on a 1-5 Likert
	scale. For CMOS, listeners are asked to compare pairs of audio generated by systems A and B and indicate which of the two audio they prefer and choose one of the following scores: 0 indicating no difference, 1 indicating small difference, 2 indicating a large difference, and 3 indicating a very large difference. For audio quality evaluation (MOS-Q and CMOS-Q), we tell listeners to \textit{"focus on examing the naturalness of audio quality (noise, timbre, sound clarity, and high-frequency details), and ignore the differences of prosody and rhythm (e.g., pitch, energy, and duration)"}. For prosody evaluations (MOS-P and CMOS-P), we tell listeners to \textit{"focus on examing the naturalness of prosody and rhythm (e.g., pitch, energy, and duration), and ignore the differences in audio quality (noise, timbre, sound clarity, and high-frequency details)"}. The screenshots of instructions for testers are shown in Fig. \ref{figure:screenshots}.
	
	\section{Additional Experiments}
	\subsection{CMOS on Variational Generator with Different Samples}
	As our SyntaSpeech follows the variational generator in PortaSpeech, we perform a case study to demonstrate that SyntaSpeech could generate more natural audio than its baseline PortaSpeech, using a variety of latent variables of VAE. To be specific, for each input text, we generate 5 audios with different input noise and compare the audio quality and prosody between PortaSpeech and SyntaSpeech.
	
	\subsection{More objective evaluations}
	In this section, we analyze the impact of each component in SyntaSpeech on the training objectives, including mel reconstruction loss (a L1 loss), duration predictor loss (an MSE loss), and KL-divergence loss of variational generator. To this end, we conduct extensive ablation studies on SyntaSpeech, and the learning curves in LJSpeech are shown in Fig. \ref{figure:disable_grad}. All curves are smoothed with a moving average with a window size of 100. It can be seen that syntactic graph encoder in duration predictor (GDP) significantly improves the convergence of duration predictor loss, and slightly improve the KL loss and mel reconstruction loss. By contrast, syntactic graph encoder in prior flow (GEF) greatly improves the KL loss and also has a small improvement to duration predictor loss and mel reconstruction loss. Besides, our SyntaSpeech achieves the best performance from the perspectives of all objective metrics. We also observe that replacing the syntactic graph (SG) with the complete graph (CG) leads to a similar performance to vanilla PortaSpeech, which proves the necessity of the introduced syntactic information.
	
	\begin{figure}[!t]
		\centering
		\subfloat[Mel Reconstruction Loss]{\includegraphics[scale=0.4]{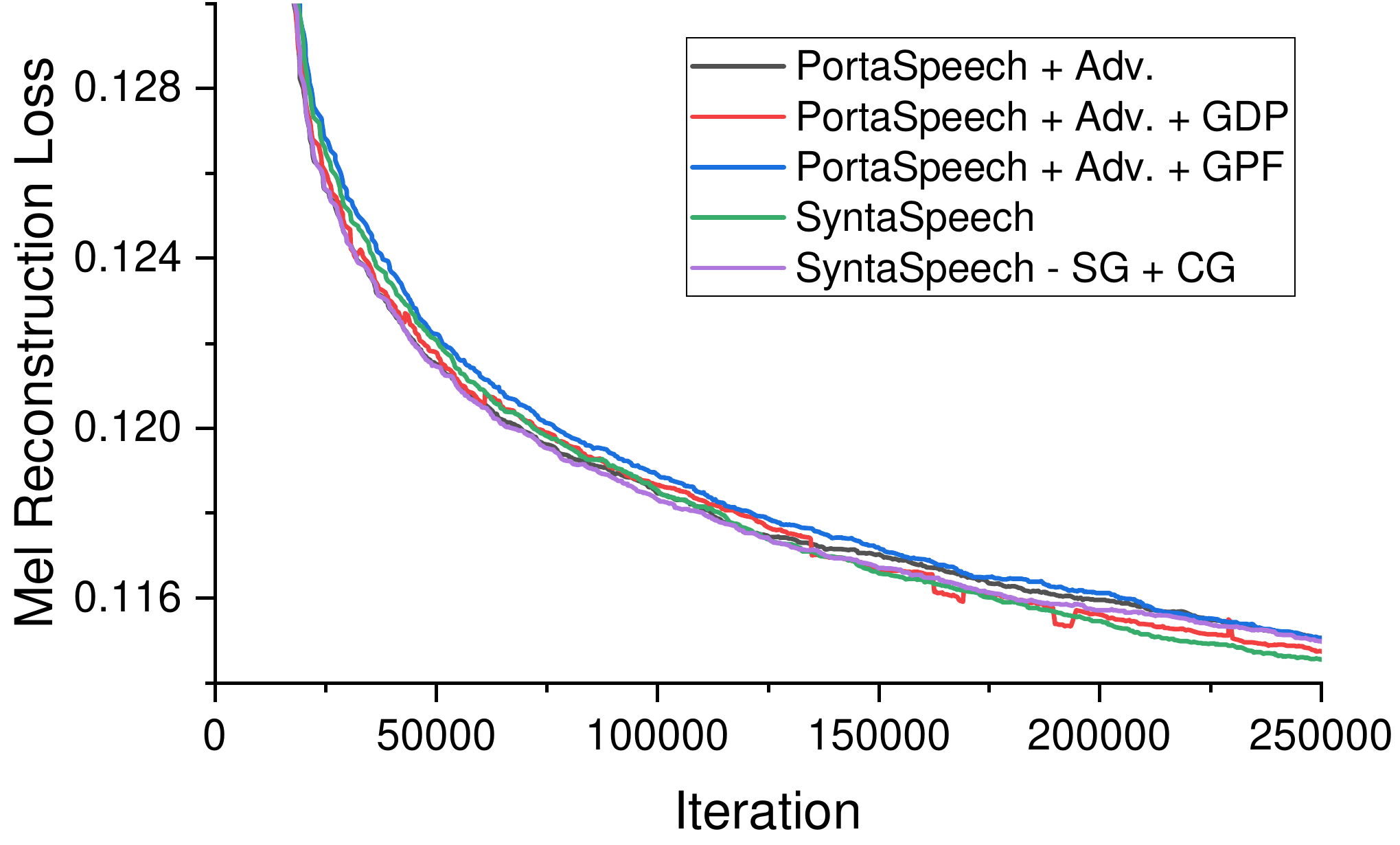}%
			\label{figure:ml_disc1}}
		\hfil
		\subfloat[Duration Predictor Loss]{\includegraphics[scale=0.4]{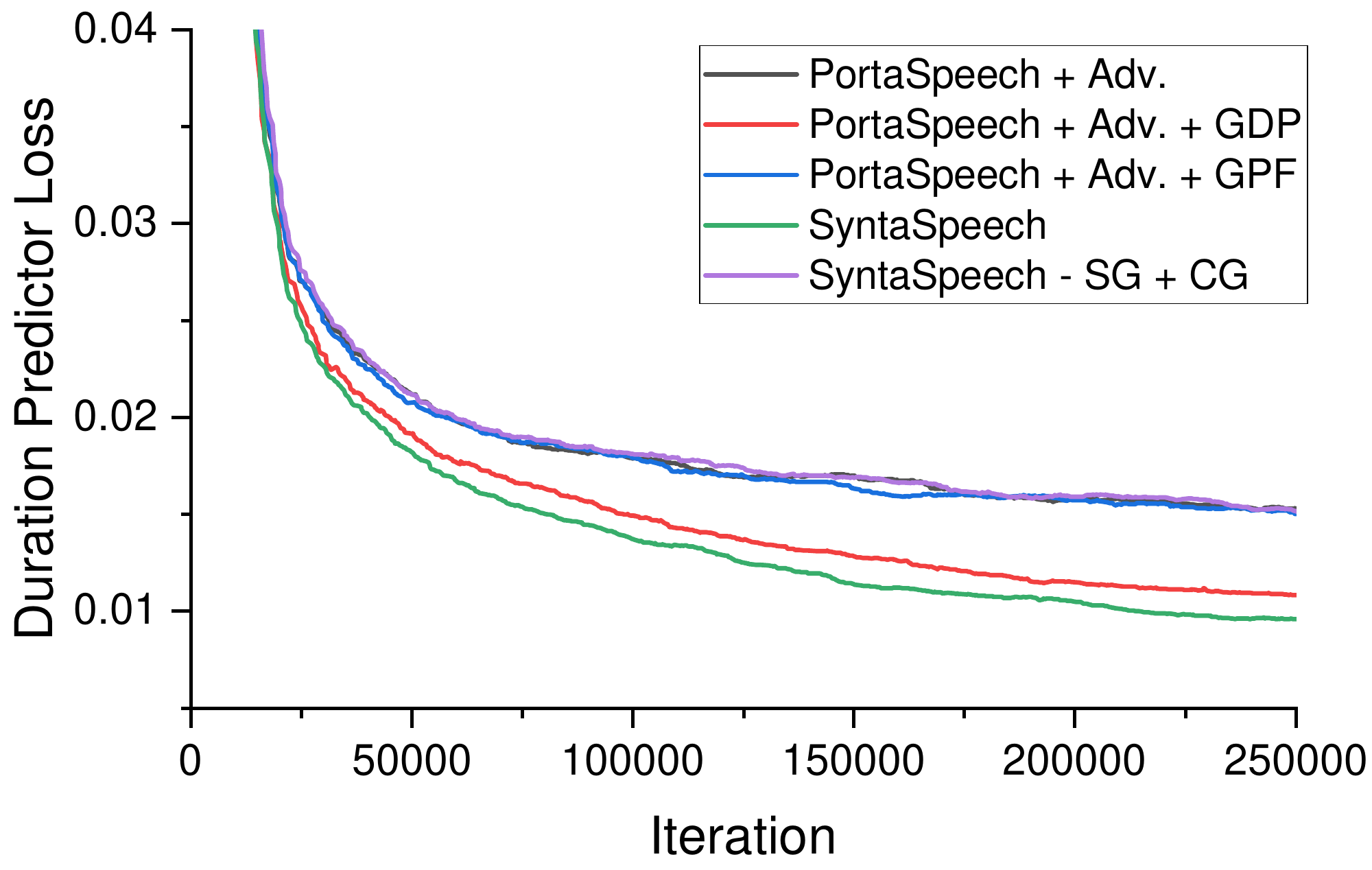}%
			\label{figure:ml_disc2}}
		\hfil
		\subfloat[KL Loss]{\includegraphics[scale=0.4]{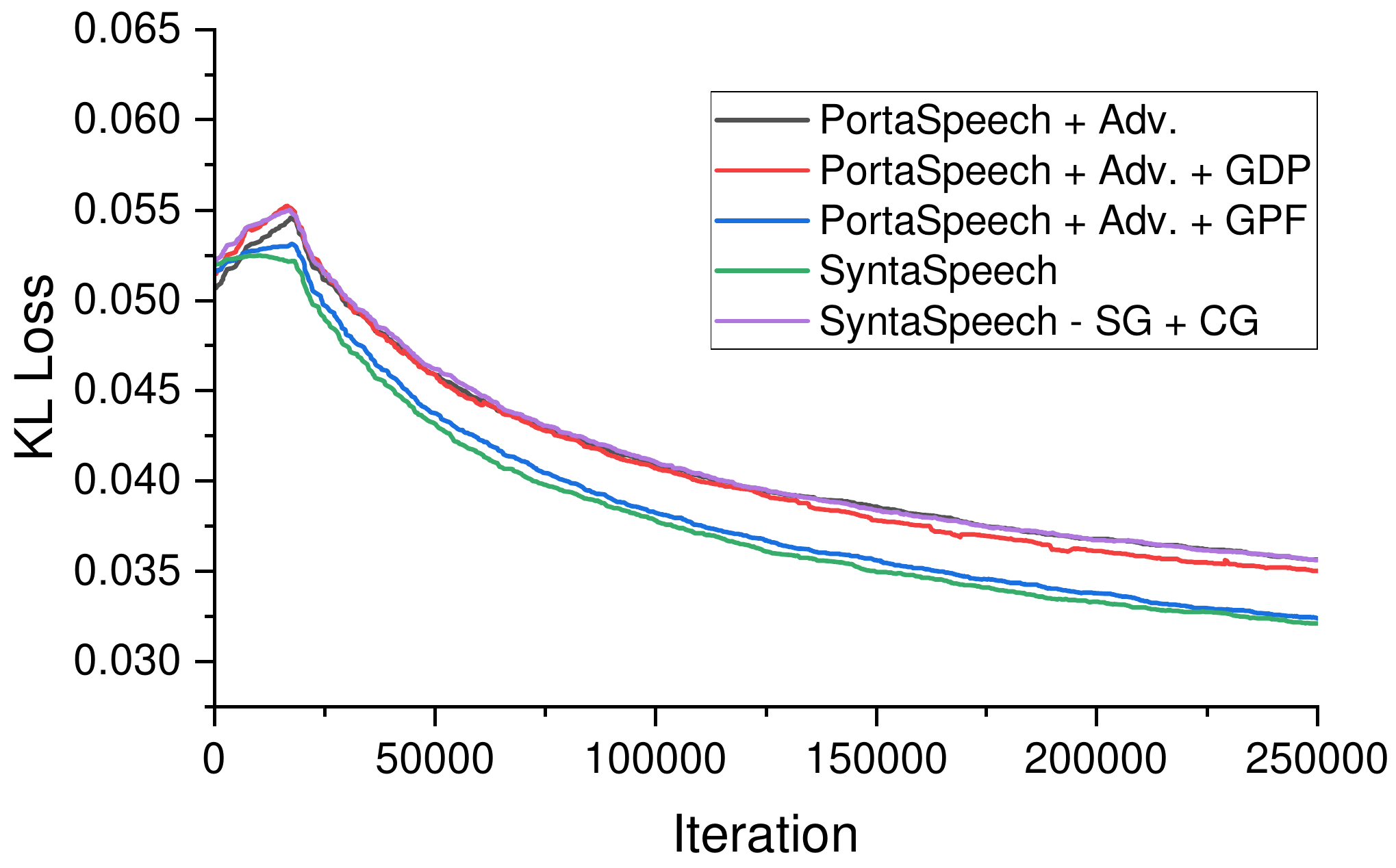}%
			\label{figure:ml_disc2}}
		\caption{The learning curves of several training losses.}
		\label{figure:disable_grad}	
	\end{figure}

	\subsection{Syntactic graph encoder's impacts on audio quality}
	We also test the syntactic graph encoder's impacts on audio quality (CMOS-Q). As shown in Tab. \ref{tab:cmos-q}, removing syntactic graph encoder from duration predictor or prior flow leads to a slightly worse audio quality.
	\begin{table}
		\centering
		\begin{tabular}{lccc}
			\toprule
			Settings  & LJSpeech & Biaobei & LibriTTS  \\
			\toprule
			\textit{SyntaSpeech}       & $0.000$  & $0.000$ & $0.000$      \\
			\toprule
			\textit{- GDP}    & $0.009$  & $-0.073$ & $-0.019$     \\
			\textit{- GPF}   & $-0.010$  & $-0.069$  & $-0.038$  \\
			\textit{- GDP - GPF}   & $-0.031$  & $-0.078$  & $-0.052$  \\
			\textit{- SG + CG} & $-0.040$ & $-0.062$  &  $-0.051$  \\
			\bottomrule
		\end{tabular}
		\caption{CMOS-Q comparisons for ablation studies. \textit{Adv} denotes multi-length adversarial training, \textit{GDP} denotes using graph encoder in duration predictor, \textit{CG} denotes using complete graph instead of the syntactic graph (\textit{SG}).}
		\label{tab:cmos-q}
	\end{table}
	
	\subsection{Adversarial Training's Impacts on Prosody}
	
	We also test the syntactic graph encoder's impacts on prosody prediction (CMOS-P). As shown in Tab. \ref{tab:cmos-p}, replacing the post-net (PN) with adversarial training (Adv) results in a better prosody score.
	
	\begin{table}
		\centering
		\begin{tabular}{lccc}
			\toprule
			Settings  & LJSpeech & Biaobei & LibriTTS  \\
			\toprule
			\textit{PortaSpeech}       & $0.000$  & $0.000$ & $0.000$      \\
			\textit{- PN + Adv.}    & $0.010$  & $0.045$ & $0.009$     \\
			\midrule
			\textit{SyntaSpeech}   & $0.000$  & $0.000$  & $0.000$  \\
			\textit{- Adv. + PN}   & $-0.018$  & $-0.117$  & $-0.020$  \\
			\bottomrule
		\end{tabular}
		\caption{CMOS-P comparisons for ablation studies. \textit{PN} denotes post-net in PortaSpeech, and \textit{Adv} means our adversarial training.}
		\label{tab:cmos-p}
	\end{table}
	
	\section{Potential Negative Societal Impacts}
	SyntaSpeech improves the prosody and audio quality of the synthesized speech voice, which may cause unemployment for people with related occupations such as broadcaster and radio host. In addition, unauthorized voice cloning or production of fake media may cause voice security issues.

\end{document}